\documentstyle[12pt,epsf]{article}
\begin{document}
\baselineskip 0.7cm

\newcommand{\gsim}{ \mathop{}_{\textstyle \sim}^{\textstyle >} }
\newcommand{\lsim}{ \mathop{}_{\textstyle \sim}^{\textstyle <} }
\newcommand{\vev}[1]{ \left\langle {#1} \right\rangle }
\newcommand{\EV}{ {\rm eV} }
\newcommand{\KEV}{ {\rm keV} }
\newcommand{\MEV}{ {\rm MeV} }
\newcommand{\GEV}{ {\rm GeV} }
\newcommand{\TEV}{ {\rm TeV} }
\newcommand{\mgut}{M_{GUT}}
\newcommand{\mint}{M_{I}}
\newcommand{\mgra}{M_{3/2}}
\newcommand{\mgy}{m_{G1}}
\newcommand{\mgl}{m_{G2}}
\newcommand{\mgc}{m_{G3}}
\newcommand{\nuR}{\nu_{R}}
\newcommand{\e}{{\rm e}}
\renewcommand{\thefootnote}{\fnsymbol{footnote}}
\setcounter{footnote}{1}

\begin{titlepage}

\begin{flushright}
UT-840
\end{flushright}

\vskip 0.35cm
\begin{center}
{\large \bf  Quark and Lepton Mass Matrices \\
             in the SO(10) Grand Unified Theory \\
             with Generation Flipping}
\vskip 1.2cm
Yasunori Nomura and Takashi Sugimoto

\vskip 0.4cm

{\it Department of Physics, University of Tokyo, 
     Tokyo 113-0033, Japan}
\vskip 1.5cm

\abstract{We investigate the SO(10) grand unified model with generation
 flipping.
 The model contains one extra matter multiplet $\psi({\bf 10})$ and it
 mixes with the usual matter multiplets $\psi_i({\bf 16})$ when the
 SO(10) is broken down to SU(5).
 We find the parameter region of the model in which the observed quark
 masses and mixings are well reproduced.
 The resulting parameter region is consistent with the observation that
 only $\psi_i({\bf 16})$ have a source of hierarchies and indicates that 
 the mixing between second and third generations tends to be large in
 the lepton sector, which is consistent with the observed maximal mixing
 of the atmospheric neutrino oscillation.
 We also show that the model can accommodate MSW and vacuum oscillation
 solutions to the solar neutrino deficit depending on the form of the
 Majorana mass matrix for the right-handed neutrinos.}

\end{center}
\end{titlepage}

\renewcommand{\thefootnote}{\arabic{footnote}}
\setcounter{footnote}{0}

%
%
%       *** Main Part ***
%
%

\section{Introduction}

The grand unified theory (GUT) \cite{GUT} is a very attractive idea
which unifies all gauge interactions of the standard model (SM) 
and explains otherwise peculiar U(1) hypercharge assignments.
In particular, its supersymmetric (SUSY) version \cite{SUSY-GUT} has
achieved great successes of gauge coupling \cite{GC-Unif} and $b-\tau$
Yukawa coupling \cite{BT-Unif} unifications, so that various SUSY GUT
models have been proposed so far.

Recently, Super-Kamiokande Collaboration has reported convincing
evidence for atmospheric neutrino oscillation with mass squared
difference 
$\delta m_{\rm atm}^2 \simeq 5 \times 10^{-3}~\EV^2$ and nearly maximal
mixing angle $\sin^2 2\theta_{\rm atm} \gsim 0.8$ \cite{SuperK}.
In the SUSY SM, neutrino masses come from the effective superpotential
\begin{eqnarray}
  W_{\rm eff} = \sum_{i,j=1}^{3} \frac{\kappa_{ij}}{M_R} l_i l_j H_u H_u,
\end{eqnarray}
where $l_i, H_u$ are SU(2)$_L$-doublet lepton chiral multiplets and the
Higgs doublet chiral multiplet giving masses for the up-type quarks,
respectively.
Here, $i,j = 1,\cdots,3$ are generation indices, $\kappa_{ij}$ denote
dimensionless coupling constants and $M_R$ is the scale at which this
operator is generated.
The observed mass squared difference $\delta m_{\rm atm}^2$ implies that
the scale $M_R$ is substantially lower than the gravitational scale 
$M_G \simeq 2.4 \times 10^{18}~\GEV$.
Thus, it is natural to think that these effective operators arise from
virtual exchange of some heavy fields of masses $M_R$ through see-saw
mechanism \cite{Seesaw}.
This motivates SO(10) GUT's \cite{SO-10}, which attain complete
unification of all quarks and leptons in a single generation together
with the right-handed neutrinos.

The minimal version of SUSY SO(10) GUT contains three generations of
quarks and leptons $\psi_i({\bf 16})$ belonging to ${\bf 16}$ of the
SO(10)$_{\rm GUT}$ and one Higgs $H({\bf 10})$.
This minimal model is known to yield a mass degeneracy of up-type and
down-type quarks and vanishing quark flavor mixing (CKM \cite{CKM}
mixing) \cite{minimal_SO-10}, so that it should obviously be extended in 
order to be reconciled with the observed pattern of quark masses and
mixings.
However, these predictions are not altogether ridiculous as a zero-th
order approximation (though not fully realistic) in the quark sector,
since both up-type and down-type quarks have hierarchical mass patterns
and the CKM mixings are all small in nature.
What requires large deviation from these predictions seems only the
large mixing observed in the atmospheric neutrino oscillation.
Thus, we would like to extend the minimal SO(10) model to be able to
accommodate the large mixing in the lepton sector, keeping their
successful approximate relations in the quark sector.
One way to achieve this extension is to slightly modify 
SU(5)-${\bf 5}^*$ components of $\psi_i({\bf 16})$
(nonparallel family structure \cite{nonpararell}).\footnote{
For other attempts to realize realistic quark and lepton mass matrices
in SO(10) GUT, see Ref.~\cite{other}.}
Since ${\bf 5}^*$ of the standard SU(5)$_{\rm GUT}$ contains
right-handed (down-type) quarks and left-handed leptons, this
modification is directly transmitted only to the lepton flavor mixing
matrix (MNS \cite{MNS} matrix).
As for the quark sector, the effect of the modification is transmitted
to the CKM matrix through a diagonalization of the down-type quark mass
matrix, inducing small CKM mixings.

In a previous paper, Yanagida and one of the authors (Y.N.) have
proposed a SO(10) GUT model which can accommodate realistic quark and
lepton mass matrices along the above line of reasoning \cite{previous}.
The model contains one extra matter multiplet $\psi({\bf 10})$ belonging 
to ${\bf 10}$ of the SO(10)$_{\rm GUT}$ and it mixes with three 
$\psi_i({\bf 16})$ when the SO(10)$_{\rm GUT}$ is broken down to SU(5)
\cite{extra_10}.
As a result, the low-energy quarks and leptons are three linear
combinations of $\psi_i({\bf 16})$ and $\psi({\bf 10})$, and realistic
masses and mixings are obtained with the right-handed neutrino Majorana
mass matrix $M_N$ proportional to the unit one.

In this paper, we extend the previous model \cite{previous} and
investigate whether it can really accommodate the observed values of
quark masses and mixings quantitatively.
We find that the model reproduces well the observed quark masses and the
CKM matrix elements.
The resulting parameter region is consistent with the observation that
only $\psi_i({\bf 16})$ have a source of hierarchies such as U(1) flavor 
symmetry charges \cite{FN}.
Furthermore, we discuss flavor mixings in the lepton sector under
certain simplifying assumptions.
We find that the model can naturally explain the observed atmospheric
neutrino deficit \cite{atm-osc, SuperK} and solar neutrino deficit 
\cite{sol-osc1, sol-osc2} by the neutrino oscillation 
\cite{nu_osci, MNS}.
The resulting mass squared differences and mixing angles depend on the
form of the Majorana mass matrix $M_N$.
We consider two cases that $M_N$ is proportional to the unit matrix
and that it has a hierarchy consistent with the above observation that
only $\psi_i({\bf 16})$ have a source of hierarchies.

This paper is organized as follows.
In section 2, we briefly review the model in Ref.~\cite{previous} and
slightly extend it removing unwanted mass degeneracy between down-type
quarks and charged leptons.
In section 3, we investigate the model and find a parameter region where
it is consistent with all the observed quark masses and mixings
including a CP-violating phase.
We discuss masses and mixings in the lepton sector, particularly
solutions to the solar neutrino deficit, in section 4.
Section 5 is devoted to our conclusions.
We also add an appendix, in which we give a detailed derivation of quark 
and lepton mass matrices in the model.

\section{The model}

In this section, we explain the model proposed in Ref.~\cite{previous}
and set up the framework for the
phenomenological analyses given in the next section.

The model has three generations of $\psi_i({\bf 16})$ and one
$\psi({\bf 10})$ as matter multiplets.
We take a basis where the original Yukawa coupling matrix of the Higgs
field $H({\bf 10})$ to the matter fields $\psi_i({\bf 16})$ is real and
diagonal:
\begin{eqnarray}
  W = \frac{1}{2} \sum_{i=1}^{3} h_i\, \psi_i({\bf 16})\, 
      \psi_i({\bf 16})\, H({\bf 10}).
\label{diagonal_Yukawa}
\end{eqnarray}
This leads to a diagonal mass matrix for the up-type quarks as
\begin{eqnarray}
  M_u = m_t \pmatrix{
          \hat{m}_u &     0     & 0  \cr
              0     & \hat{m}_c & 0  \cr
              0     &     0     & 1 \cr},
\end{eqnarray}
where $\hat{m}_u$ and $\hat{m}_c$ are defined as 
\begin{eqnarray}
  \hat{m}_u = m_u / m_t, \qquad
  \hat{m}_c = m_c / m_t,
\end{eqnarray}
and $m_u, m_c$ and $m_t$ are given by
\begin{eqnarray}
  m_u = h_1\, \vev{{\bf 5}_H}, \qquad
  m_c = h_2\, \vev{{\bf 5}_H}, \qquad
  m_t = h_3\, \vev{{\bf 5}_H}.
\end{eqnarray}
Here, ${\bf 5}_H$ is a SU(5)-${\bf 5}$ component of 
$H({\bf 10})$.
At this stage, the down-type quark mass matrix $M_{d/l}$ and neutrino
Dirac mass matrix $M_{\nu D}$ are completely the same with $M_u$ except
for the difference between the vacuum expectation values of the Higgs
fields.

We now assume that the SO(10)$_{\rm GUT}$ is broken down to 
SU(5) by condensation of Higgs 
$\vev{\chi({\bf 16})} = \vev{\bar{\chi}({\bf 16}^*)} = V$ with $V$ being 
$\sim 10^{16}~\GEV$.
This GUT breaking also induces a mass term for the matter multiplets
through the following superpotential:
\begin{eqnarray}
  W = \sum_{i=1}^{3} f_i \e^{i\eta_i}\, 
      \psi_i({\bf 16})\, \psi({\bf 10}) \vev{\chi({\bf 16})}.
\label{mass_matter}
\end{eqnarray}
Namely, a linear combination $\tilde{\bf 5}^* \propto
\sum_{i=1}^{3} f_i \e^{i\eta_i}\, {{\bf 5}^*_i}'$ in $\psi_i({\bf 16})$
receives a GUT scale mass together with ${\bf 5}_{\psi}$ in 
$\psi({\bf 10})$, and the other two linear combinations, ${\bf 5}^*_1$
and ${\bf 5}^*_3$, in $\psi_i({\bf 16})$ remain as massless fields.
The relation between these fields can be parametrized as
\begin{eqnarray}
  \pmatrix{{{\bf 5}^*_1}' \cr
           {{\bf 5}^*_2}' \cr
           {{\bf 5}^*_3}' \cr}
  = 
  \pmatrix{
          \e^{-i\delta_1-i\delta_2} & 0 &        0        \cr
          0 & \e^{-i\delta_1+i\delta_2} &        0        \cr
                0       &       0       & \e^{2i\delta_1} \cr}
  \pmatrix{
          \cos\alpha_1 & -\sin\alpha_1 & 0 \cr
          \sin\alpha_1 &  \cos\alpha_1 & 0 \cr
                0      &        0      & 1 \cr} \nonumber\\
  \times
  \pmatrix{
                   \cos\alpha_2       & 0 & \e^{-i\delta_3}\sin\alpha_2 \cr
                        0             & 1 &               0             \cr
          -\e^{i\delta_3}\sin\alpha_2 & 0 &          \cos\alpha_2       \cr}
  \pmatrix{\tilde{\bf 5}^* \cr
           {\bf 5}^*_1     \cr
           {\bf 5}^*_3     \cr},
\end{eqnarray}
without a loss of generality (see appendix).
Here, $\alpha_1, \alpha_2$ and $\delta_i$ are functions of $f_i$ and
$\eta_i$.
Thus, the low-energy quark and lepton fields belonging to ${\bf 5}^*$ of 
SU(5) are ${\bf 5}^*_1$, ${\bf 5}^*_3$ and SU(5)-${\bf 5}^*$ of
$\psi({\bf 10})$ ({\it i.e.} ${\bf 5}^*_{\psi}$), so that the relations
$M_u = M_{\nu D} \propto M_{d/l}$ are removed.

The down-type quark mass matrix is, however, incomplete, since 
${\bf 5}^*_{\psi}$ does not have any Yukawa coupling to $H({\bf 10})$.
To solve this problem we introduce a pair of Higgs $H({\bf 16})$ and 
$\bar{H}({\bf 16}^*)$ and consider a superpotential\footnote{
We can take $\lambda$ and $M$ to be real using phase degrees of freedom
in $H({\bf 16})$ and $\bar{H}({\bf 16}^*)$.
Strictly speaking, $\phi_i$ in Eq.~(\ref{down-type_giving}) are defined
in this basis.}
\begin{eqnarray}
  W = \lambda\, H({\bf 10})\, \bar{H}({\bf 16}^*)\, \bar{\chi}({\bf 16}^*) 
      + M\, H({\bf 16})\, \bar{H}({\bf 16}^*).
\label{Higgs_pot}
\end{eqnarray}
U(1) $R$-symmetry may be useful to have this form of superpotential.
The U(1)$_R$ charges are given in Table~\ref{R_charge}.
\begin{table}
\begin{center}
\begin{tabular}{|c|ccccccc|}  \hline 
  & $H({\bf 10})$ & $H({\bf 16})$ & $\bar{H}({\bf 16}^*)$ & 
    $\chi({\bf 16})$ & $\bar{\chi}({\bf 16}^*)$ & $\psi_i({\bf 16})$ &
    $\psi({\bf 10})$ \\ \hline
  $R$ & 0 & 0 & 2 & 0 & 0 & 1 & 1 \\ \hline
\end{tabular}
\end{center}
\caption{U(1)$_R$ charges.}
\label{R_charge}
\end{table}
The GUT condensation $\vev{\bar{\chi}({\bf 16}^*)} \neq 0$
induces a mass mixing between ${\bf 5}^*$'s of $H({\bf 10})$ and 
$H({\bf 16})$ ({\it i.e.} ${\bf 5}^*_{H({\bf 10})}$ and 
${\bf 5}^*_{H({\bf 16})}$).
Thus, a linear combination
\begin{eqnarray}
  {\bf 5}^*_H &=& \cos\theta\, {\bf 5}^*_{H({\bf 10})}
                      + \sin\theta\, {\bf 5}^*_{H({\bf 16})},\\
  \tan\theta &=& - \frac{\lambda\, \vev{\bar{\chi}({\bf 16}^*)} }{M},
\end{eqnarray}
remains as a massless Higgs in the standard SU(5)$_{\rm GUT}$ and
contributes to the quark and lepton mass matrices.
Then, ${\bf 5}^*_H$ can couple to ${\bf 5}^*_{\psi}$ through the
superpotential 
\begin{eqnarray}
  W = \sum_{i=1}^{3} g_i \e^{i\phi_i}\, 
      \psi_i({\bf 16})\, \psi({\bf 10})\, H({\bf 16}),
\label{down-type_giving}
\end{eqnarray}
and the down-type quark and charged-lepton mass matrix is given by
\begin{eqnarray}
  M_{d/l} = m_t \frac{\cos\theta}{\tan\beta} \pmatrix{
          -\hat{m}_u\sin\alpha_1 & x\e^{i\phi_x} & 
               \hat{m}_u\cos\alpha_1\sin\alpha_2 \cr
           \hat{m}_c\cos\alpha_1 & y\e^{i\phi_y} & 
               \hat{m}_c\sin\alpha_1\sin\alpha_2 \cr
                        0        &       z       & 
                          \cos\alpha_2           \cr}.
\label{M_d-l}
\end{eqnarray}
Here, $\tan\beta \equiv \vev{{\bf 5}_H} / \vev{{\bf 5}^*_H}$ and
\begin{eqnarray}
  x = \frac{g_1}{h_{3}}\tan\theta, \qquad 
  y = \frac{g_2}{h_{3}}\tan\theta, \qquad
  z = \frac{g_3}{h_{3}}\tan\theta,
\end{eqnarray}
\begin{eqnarray}
  \phi_x = \phi_1 - \phi_3 + 3\delta_1 + \delta_2 + \delta_3, \qquad
  \phi_y = \phi_2 - \phi_3 + 3\delta_1 - \delta_2 + \delta_3,
\end{eqnarray}
(for a detailed derivation, see appendix).
Note that in the present model $\tan\beta$ can take a value in a wide
range due to the presence of $\cos\theta$ in Eq.~(\ref{M_d-l}).

The Dirac mass matrix $M_{\nu D}$ for neutrinos is also incomplete,
since ${\bf 5}^*_{\psi}$ never couples to SU(5)-singlets of 
$\psi_i({\bf 16})$ ({\it i.e.} ${\bf 1}_i$).
However, the following nonrenormalizable interactions give desired
couplings:
\begin{eqnarray}
  W = \sum_{i=1}^{3} k_i \e^{i\varphi_i}\, 
      \psi_i({\bf 16})\, \psi({\bf 10})\, H({\bf 10}) 
      \frac{\bar{\chi}({\bf 16}^*)}{M_G}.
\label{nonren_nuD}
\end{eqnarray}
Together with the original couplings in Eq.~(\ref{diagonal_Yukawa}), 
the nonrenormalizable interactions Eq.~(\ref{nonren_nuD}) yield
\begin{eqnarray}
  M_{\nu D} = m_t \pmatrix{
            -\hat{m}_u\sin\alpha_1 & 
              \delta_x\e^{i\varphi_x} & 
              \hat{m}_u\e^{i\varphi_u}\cos\alpha_1\sin\alpha_2 \cr
            \hat{m}_c\e^{i\varphi_t}\cos\alpha_1 & 
              \delta_y\e^{i\varphi_y} & 
              \hat{m}_c\e^{i\varphi_v}\sin\alpha_1\sin\alpha_2 \cr
            0 & 
              \delta_z\e^{i\varphi_z} & 
              \e^{i\varphi_w}\cos\alpha_2  \cr}.
\end{eqnarray}
Here,
\begin{eqnarray}
  \delta_x = \frac{k_1}{h_{3}}\frac{\vev{\bar{\chi}({\bf 16}^*)}}{M_G}, \qquad 
  \delta_y = \frac{k_2}{h_{3}}\frac{\vev{\bar{\chi}({\bf 16}^*)}}{M_G}, \qquad
  \delta_z = \frac{k_3}{h_{3}}\frac{\vev{\bar{\chi}({\bf 16}^*)}}{M_G},
\label{neutrino_delta}
\end{eqnarray}
\begin{eqnarray}
  \matrix{
  \varphi_x = \varphi_1 + \delta_1 - \delta_2, \qquad
  \varphi_y = \varphi_2 + \delta_1 - \delta_2, \qquad
  \varphi_z = \varphi_3 + \delta_1 - \delta_2, \cr
  \varphi_t =   2\delta_2, \qquad
  \varphi_u = - 3\delta_1 - \delta_2 - 2\delta_3, \cr
  \varphi_v = - 3\delta_1 + \delta_2 - 2\delta_3, \qquad
  \varphi_w = -  \delta_3, \cr}
\end{eqnarray}
(see appendix).
Notice that $\delta_x, \delta_y, \delta_z \simeq O(10^{-2})$ as long as
$k_i/h_3 = O(1)$.
The Majorana masses for the right-handed neutrinos ${\bf 1}_i$ are given
by the following nonrenormalizable superpotential:
\begin{eqnarray}
  W = \frac{1}{2} \sum_{i,j=1}^{3} j_{ij}\, 
      \psi_i({\bf 16})\, \psi_j({\bf 16})\, 
      \frac{\bar{\chi}({\bf 16}^*)\, 
      \bar{\chi}({\bf 16}^*)}{M_G}.
\label{Majorana_mass}
\end{eqnarray}
After the SO(10)$_{\rm GUT}$ breaking 
we obtain the Majorana mass matrix for the right-handed neutrinos,
\begin{eqnarray}
  (M_N)_{ij} = \frac{j_{ij}}{|j_{33}|}\, M_R,
\end{eqnarray}
where $M_R = |j_{33}|\, V^2 / M_G$.
Then, the mass matrix $M_{\nu}$ for the light neutrinos is given by 
$M_{\nu} = M_{\nu D}^{\rm T} M_N^{-1} M_{\nu D}$ through see-saw
mechanism \cite{Seesaw}.

Now, we have SU(5)-invariant mass matrices $M_u, M_{d/l}$ and $M_{\nu}$.
However, they yield wrong SU(5)$_{\rm GUT}$ mass relations, 
$m_{\mu} = m_s$ and $m_e = m_d$, so that we have to remove these
unwanted mass relations in order to obtain realistic quark and lepton
masses.
It can be done by introducing SU(5) breaking effects into $M_{d/l}$
\cite{G-J}.
We assume that the SU(5) is broken down to the SM gauge group by a Higgs
multiplet $\Sigma({\bf 45})$ belonging to ${\bf 45}$ of the 
SO(10)$_{\rm GUT}$. 
Then, if $\Sigma({\bf 45})$ has nonrenormalizable interactions to the
matter and Higgs fields, the SU(5) breaking effect of order 
$\vev{\Sigma({\bf 45})}/M_G$ can be transmitted to the quark and lepton
mass matrices \cite{GUT-rel}.
For this purpose, we introduce the following nonrenormalizable
superpotential:
\begin{eqnarray}
  W = \sum_{i=1}^{3} g'_i \e^{i\phi'_i}\, 
      \psi_i({\bf 16})\, \psi({\bf 10})\, H({\bf 16})\, 
      \frac{\vev{\Sigma({\bf 45})}}{M_G}.
\label{GJ-superpot}
\end{eqnarray}
We consider that only $(1,2)$ and $(2,2)$ components of the down-type
quark mass matrix $M_d$ and the charged-lepton mass matrix $M_l$ are
modified from those of $M_{d/l}$ for simplicity, since the value of $z$
which gives realistic masses and mixings is relatively larger than those
of $x$ and $y$ as we shall see later.
We represent these modified components with the subscript $d$ and $l$ 
({\it i.e.} $x_d, y_d, \phi_{xd}, \phi_{yd}, x_l, y_l, \phi_{xl}$ and
$\phi_{yl}$).

With the mass matrices $M_u, M_d, M_l$ and $M_{\nu}$, we can reproduce
well the observed quark and lepton flavor structure.
In the next section, we search a parameter region of $M_u$ and $M_d$
which gives the observed quark masses and mixings including a
CP-violating phase.

\section{The masses and mixings in the quark sector}

The quark masses are estimated using various methods such as QCD sum
rules.
However, the estimated quark masses still have some uncertainties.
In our analysis, we adopt the running quark and lepton masses evaluated
at the energy of $Z$-boson mass given in Table~\ref{quark-lepton_mass}
\cite{PD-mass, GUT-mass, HO-mass} to constrain the mass matrices derived 
in section 2.
\begin{table}
\begin{center}
\begin{tabular}{|c|c|c|}  \hline 
             & $\mu = m_Z$          & $\mu = M_{\rm GUT}$  \\ \hline
  $m_u(\mu)$ & $2.2 \pm 0.7~\MEV$   & $0.98 \pm 0.31~\MEV$ \\
  $m_c(\mu)$ & $626 \pm 106~\MEV$   & $279 \pm 47~\MEV$    \\
  $m_t(\mu)$ & $175 \pm 6~\GEV$     & $110 \pm 19~\GEV$    \\ \hline
  $m_d(\mu)$ & $4.1 \pm 1.1~\MEV$   & $1.2 \pm 0.3~\MEV$   \\
  $m_s(\mu)$ & $85 \pm 19~\MEV$     & $24 \pm 5~\MEV$      \\
  $m_b(\mu)$ & $3.02 \pm 0.19~\GEV$ & $1.01 \pm 0.06~\GEV$ \\ \hline
  $m_e(\mu)$ & $0.487~\MEV$         & $0.325~\MEV$         \\
  $m_{\mu}(\mu)$  & $102.7~\MEV$    & $68.6~\MEV$          \\
  $m_{\tau}(\mu)$ & $1.747~\GEV$    & $1.171~\GEV$         \\ \hline
\end{tabular}
\end{center}
\caption{The running quark and lepton masses evaluated at the energy of
 $Z$-boson mass $m_Z$ \cite{PD-mass, GUT-mass, HO-mass} and the GUT
 scale $M_{\rm GUT} \simeq 2 \times 10^{16}~\GEV$ \cite{GUT-mass}.}
\label{quark-lepton_mass}
\end{table}

The CKM mixing angles are specified by three real parameters 
$(V_{CKM})_{us}, (V_{CKM})_{cb}$ and $(V_{CKM})_{ub}/(V_{CKM})_{cb}$
whose observed values are given in Table~\ref{CKM_parameters}
\cite{PD-mass, CP-exp}.
\begin{table}
\begin{center}
\begin{tabular}{|c|c|c|}  \hline 
                        & $\mu = m_Z$        & $\mu = M_{\rm GUT}$ \\ \hline
  $(V_{CKM})_{us}(\mu)$ & $0.215 \sim 0.224$ & $0.215 \sim 0.224$  \\
  $(V_{CKM})_{cb}(\mu)$ & $0.036 \sim 0.043$ & $0.031 \sim 0.037$  \\
  $(V_{CKM})_{ub}/(V_{CKM})_{cb}(\mu)$ & 
        $0.060 \sim 0.12$ & $0.061 \sim 0.12$ \\ \hline
  $J_{CKM}(\mu)$        & 
        $(1.5 \sim 4.4) \times 10^{-5}$ & 
        $(1.1 \sim 3.3) \times 10^{-5}$ \\ \hline
\end{tabular}
\end{center}
\caption{The observed CKM parameters (at $\mu = m_Z$) 
 \cite{PD-mass, CP-exp} and their RG-evolved values at the GUT scale 
 $M_{\rm GUT} \simeq 2 \times 10^{16}~\GEV$ \cite{GUT-mass}.}
\label{CKM_parameters}
\end{table}
CP violation effects are parametrized by the quantity called Jarlskog
parameter $J_{CKM}$ \cite{Jarlskog} defined by $J_{CKM} \equiv 
{\rm Im}\, \{(V_{CKM})_{ud}(V_{CKM})_{td}^*(V_{CKM})_{tb}
(V_{CKM})_{ub}^*\}$, which is independent of phase conventions of the SM 
fields.
The value of $J_{CKM}$ obtained from neutral meson experiments is given
in Table~\ref{CKM_parameters} \cite{CP-exp}.
In the following, we search a parameter region of the model consistent
with these observed masses and mixings.

To compare with the mass matrices obtained in section 2, we have to use
renormalization group (RG)-evolved values of the masses and mixings at
the GUT scale.
We assume particle content of the minimal SUSY SM between the weak scale 
and the GUT scale.
Then, the GUT scale values are not sensitive to the precise mass
spectrum of the SUSY particles and also to $\tan\beta$ as long as it has
a moderate value $5 \lsim \tan\beta \lsim 30$ \cite{RG-evolv}.
Thus, we fix a value $\tan\beta = 10$ for simplicity.
The obtained masses and mixings at the GUT scale using two-loop RG
equations for the Yukawa couplings are given in
Table~\ref{quark-lepton_mass} and \ref{CKM_parameters} \cite{GUT-mass}.
We will use these values to constrain the parameter space of the
model.\footnote{The GUT scale value of $m_t$ is highly dependent on the
input value of $m_t(m_Z)$, especially in the case that 
$m_t(m_Z) \gsim 180~\GEV$.
As we shall see, however, our qualitative conclusions hardly depend on
the value of $m_t(M_{GUT})$.}

Now, we search values of 
$x_d, y_d, z, \alpha_1, \alpha_2, \phi_{xd}, \phi_{yd}$ which reproduce 
down-type quark mass ratios in Table~\ref{quark-lepton_mass}, 
$32.8 < m_b/m_s < 56.3$ and $12.7 < m_s/m_d < 32.2$, and all the CKM
parameters given in Table~\ref{CKM_parameters}.
We take $\hat{m}_c$ to be $\hat{m}_c^{-1}(M_{\rm GUT}) = 394$ 
(central value) and $279$ (lowest value) in order to show the dependence 
of the results on up-type quark masses at the GUT scale.
The results hardly depend on $\hat{m}_u$ due to its smallness, so
that we fix $\hat{m}_u^{-1}(M_{\rm GUT})$ to be its central value
$112244$.

\begin{figure}
\epsfxsize=10cm
\centerline{\epsfbox{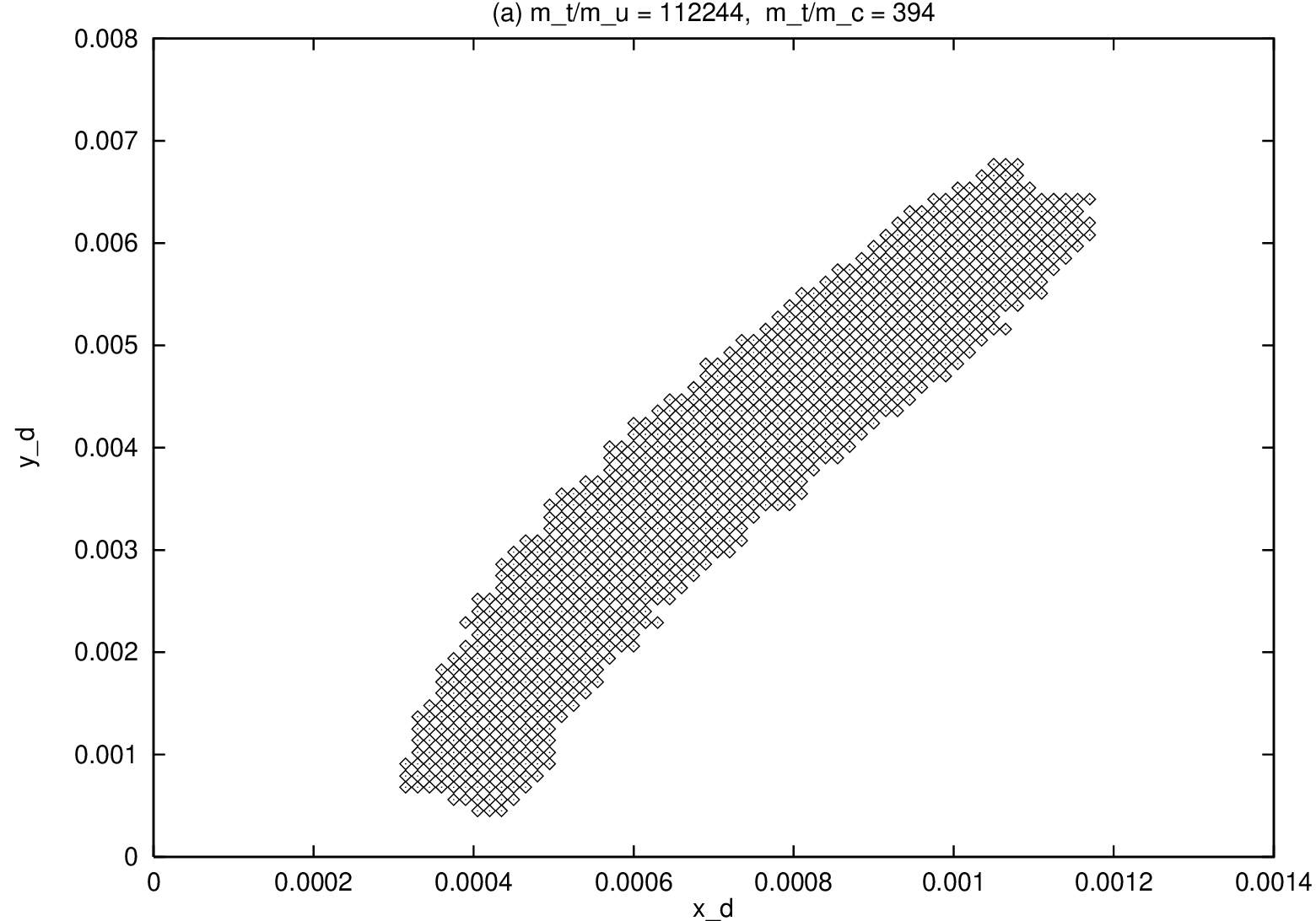}}
\vspace{1cm}
\epsfxsize=10cm
\centerline{\epsfbox{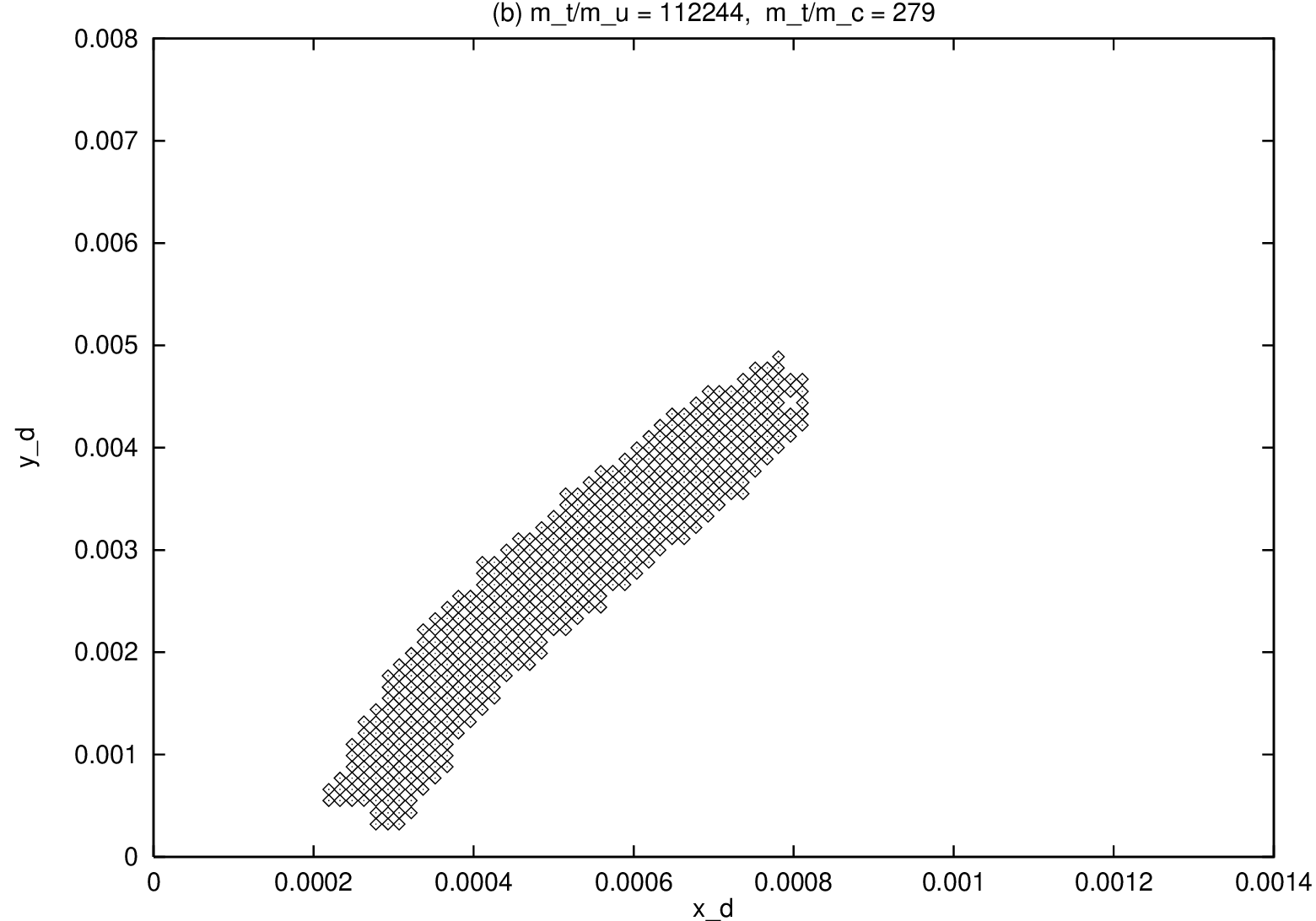}}
\caption{The parameter region of $x_d$ and $y_d$ where all quark masses
 and the CKM parameters are reproduced.
 We have set $\hat{m}_u^{-1}$ to be $112244$ and $\hat{m}_c^{-1}$ to be
 $(a) 394$ (central value) $(b) 279$ (lowest value).}
\label{fig-x_d-y_d}
\end{figure}

We find a parameter space which reproduces all quark masses and mixings 
including a CP-violating phase.
The parameter region of $x_d$ and $y_d$ is shown in
Fig.~\ref{fig-x_d-y_d}.
As is readily seen, the slope of this $x_d$-$y_d$ graph roughly gives
the reciprocal of the Cabibbo angle $\sim (0.22)^{-1}$.
Since ratios of $x_d, y_d$ and $z$ are related to the CKM mixing angles,
we find that $x_d, y_d$ and $z$ also have hierarchical structure.
Indeed, the region of $z$ is around $0.02 < z < 0.12$.
The required hierarchy is milder than that of $h_1, h_2$ and $h_3$, and
can be roughly parametrized as
\begin{eqnarray}
\begin{array}{ccccc}
                      && h_1 : h_2 : h_3 &\simeq& 
                         \epsilon^4 : \epsilon^2 : 1, \\ \\
  x_d : y_d : z &\simeq& g_1 : g_2 : g_3 &\simeq& 
                         \epsilon^2 : \epsilon : 1,
\end{array}
\label{hierarchy_1}
\end{eqnarray}
with $\epsilon = O(0.1)$.
This may indicate that only $\psi_i({\bf 16})$ have a source of
hierarchies such as U(1) flavor symmetry charges 
(see Eqs.~(\ref{diagonal_Yukawa}, \ref{down-type_giving})).
To reproduce the observed CP violation (the value of $J_{CKM}$
given in Table~\ref{CKM_parameters}) $\phi_{yd}$ has to be around
$3\pi/2$ ($10\pi/8 \lsim \phi_{yd} \lsim 15\pi/8$ in both cases of
$\hat{m}_c^{-1} = 394$ and $\hat{m}_c^{-1} = 279$).
The results hardly depend on $\phi_{xd}$, since it can be transfered
into the phase of $\hat{m}_u$ by an appropriate phase rotation of the
quark doublet $(u, d)$ and $\hat{m}_u$ is very small.

\begin{figure}
\epsfxsize=10cm
\centerline{\epsfbox{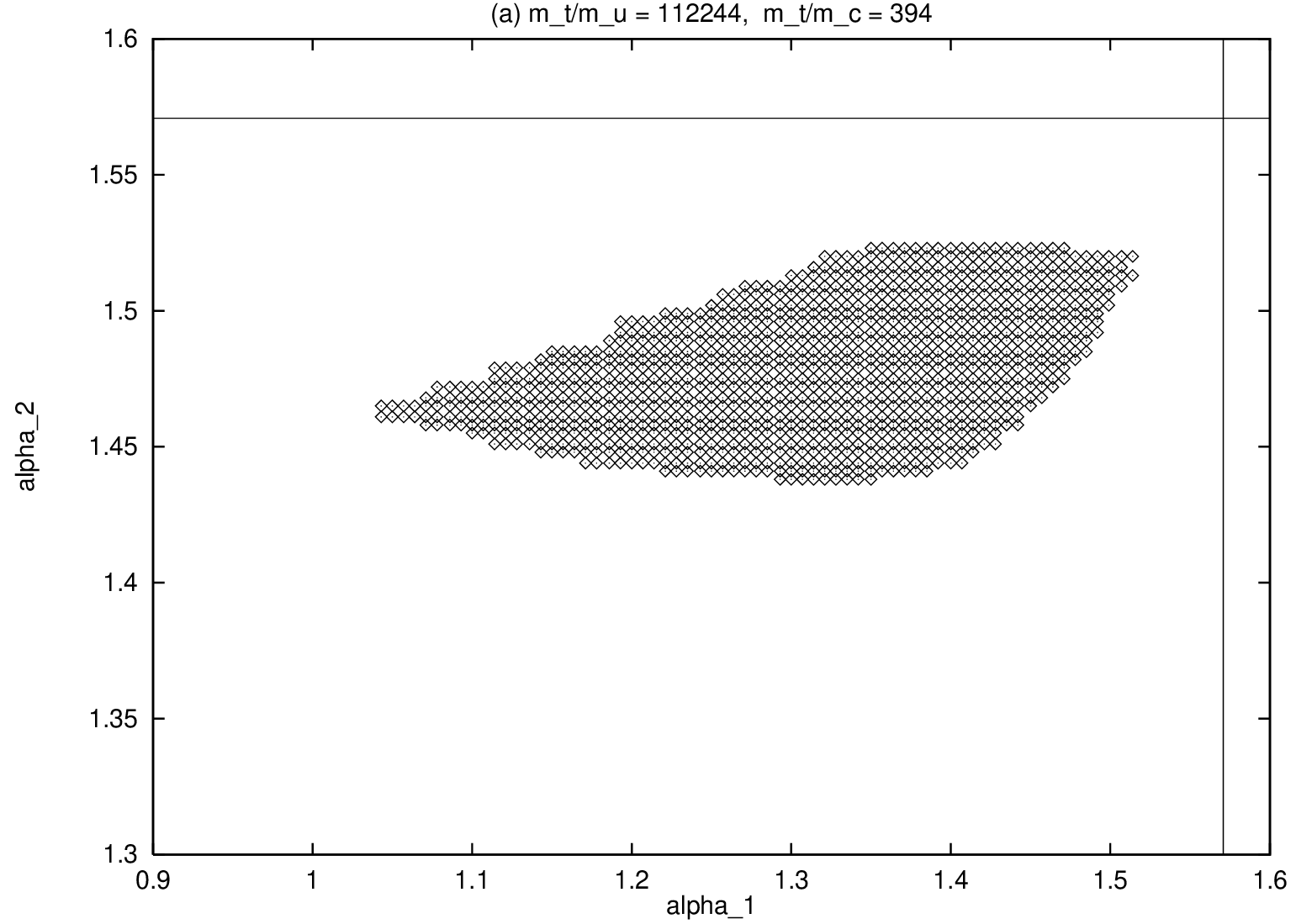}}
\vspace{1cm}
\epsfxsize=10cm
\centerline{\epsfbox{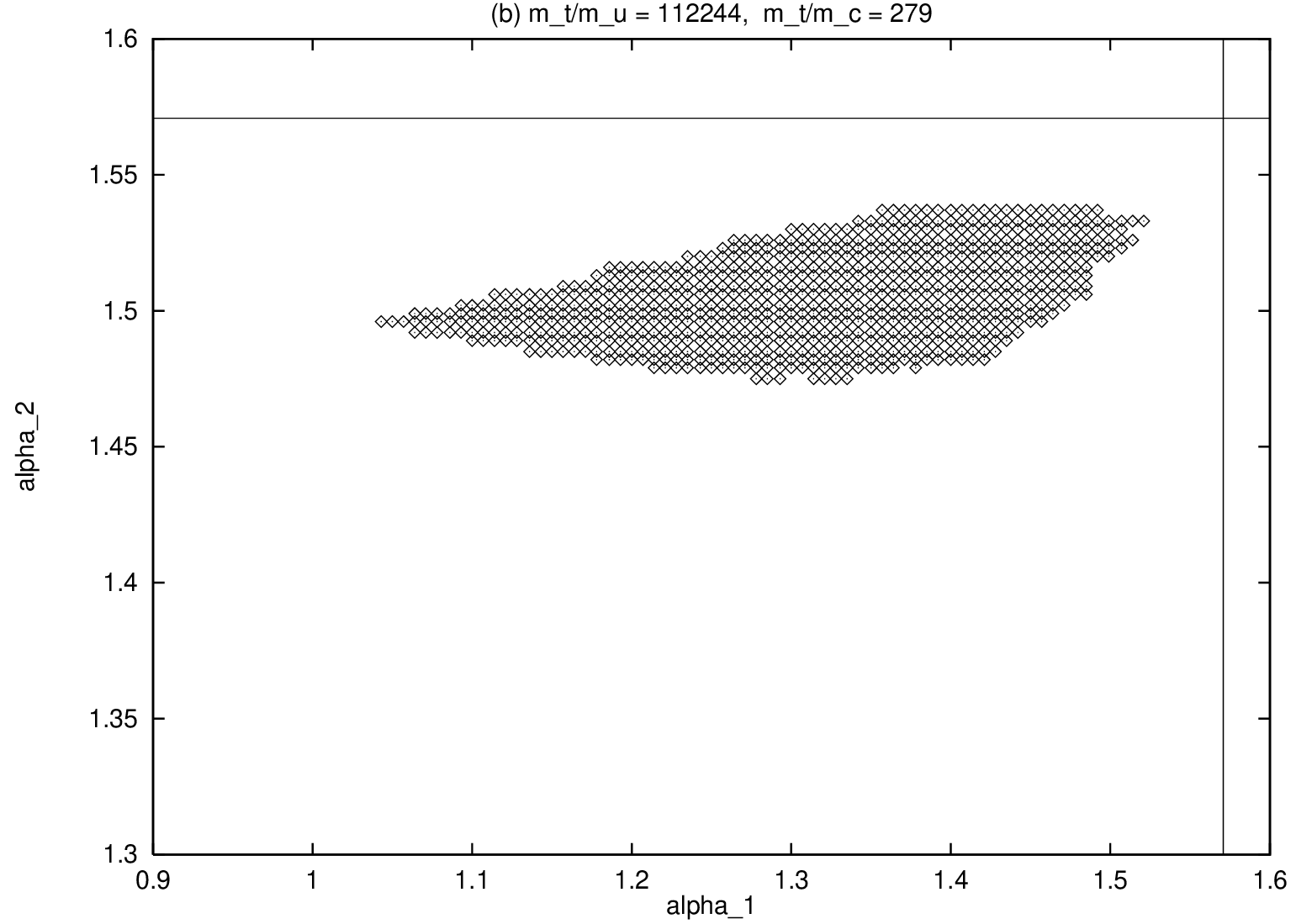}}
\caption{The parameter region of $\alpha_1$ and $\alpha_2$ where all
 quark masses and the CKM parameters are reproduced.
 $\alpha_1$ and $\alpha_2$ are represented with radians and their
 domains are $0 \leq \alpha_1 \leq \pi/2$ and $0 \leq \alpha_2 \leq \pi/2$.
 Solid lines denote $\alpha_1 = \pi/2$ and $\alpha_2 = \pi/2$.
 We have set $\hat{m}_u^{-1}$ to be $112244$ and $\hat{m}_c^{-1}$ to be
 $(a) 394$ (central value) $(b) 279$ (lowest value).}
\label{fig-a_1-a_2}
\end{figure}

The parameter region of $\alpha_1$ and $\alpha_2$ 
($0 \leq \alpha_1, \alpha_2 \leq \pi/2$) is shown in
Fig.~\ref{fig-a_1-a_2}. 
We find that both $\alpha_1$ and $\alpha_2$ are close to $\pi/2$ 
($\cos\alpha_1 \sim \cos\alpha_2 = O(0.1)$).
This implies that the extra matter multiplet $\psi({\bf 10})$ is
dominantly coupled with $\psi_3({\bf 16})$ in the superpotential 
Eq.~(\ref{mass_matter}).\footnote{
If some components of the down-type quark mass matrix other than (1,2)
and (2,2) ones are modified by nonrenormalizable interactions, the
parameter region of $(\alpha_1, \alpha_2)$ is not necessarily very close
to $(\pi/2, \pi/2)$.
This possibility will be considered elsewhere.}
The hierarchy of the couplings $f_i$ is roughly
\begin{eqnarray}
\begin{array}{ccccc}
  \cos\alpha_1\cos\alpha_2 : \sin\alpha_1\cos\alpha_2 : \sin\alpha_2 
       &=& f_1 : f_2 : f_3 &\simeq& \epsilon^2 : \epsilon : 1,
\end{array}
\label{hierarchy_2}
\end{eqnarray}
which also is consistent with the above observation that only 
$\psi_i({\bf 16})$ have a source of hierarchies 
(see Eq.~(\ref{mass_matter})).

\begin{figure}
\epsfxsize=10cm
\centerline{\epsfbox{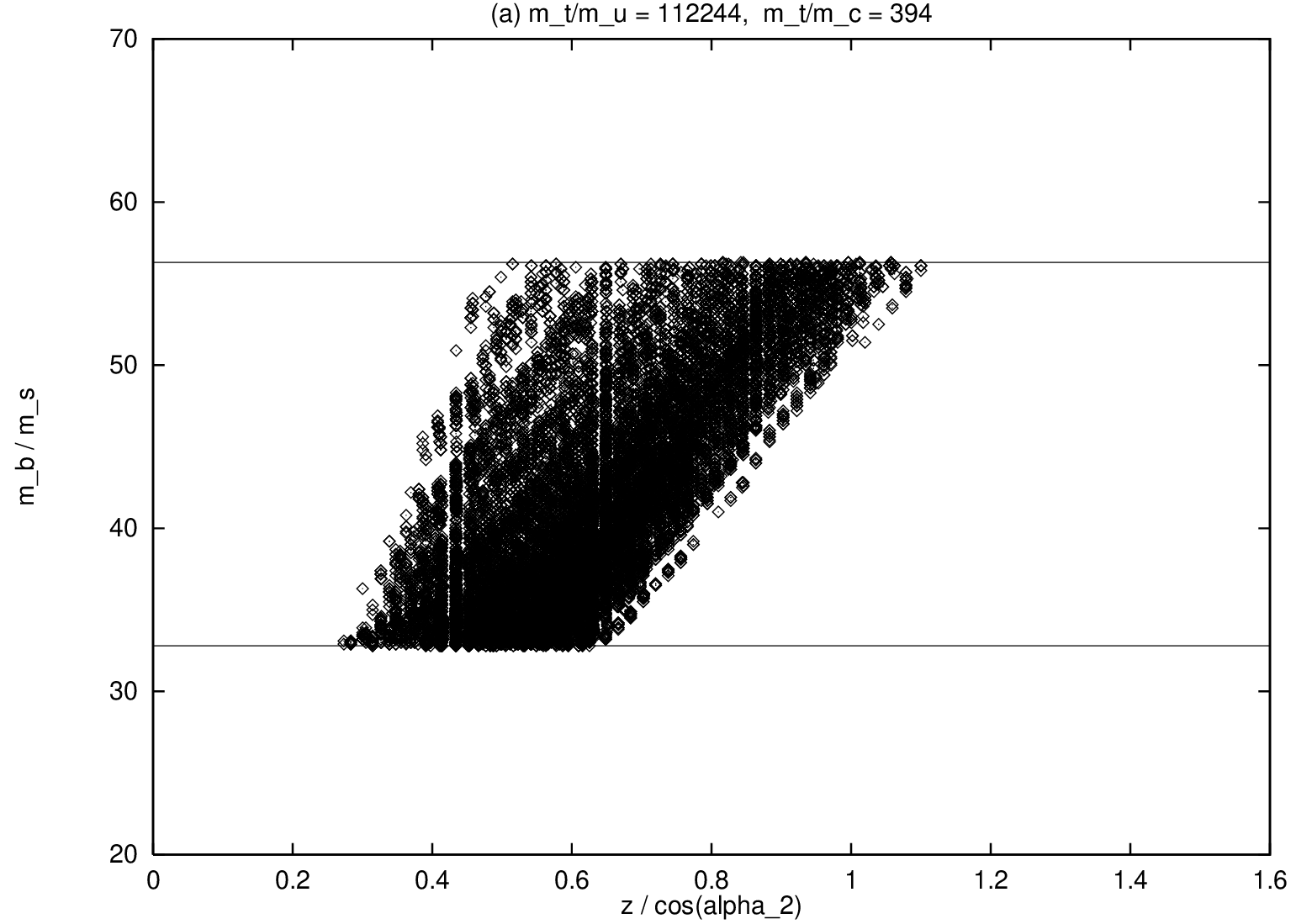}}
\vspace{1cm}
\epsfxsize=10cm
\centerline{\epsfbox{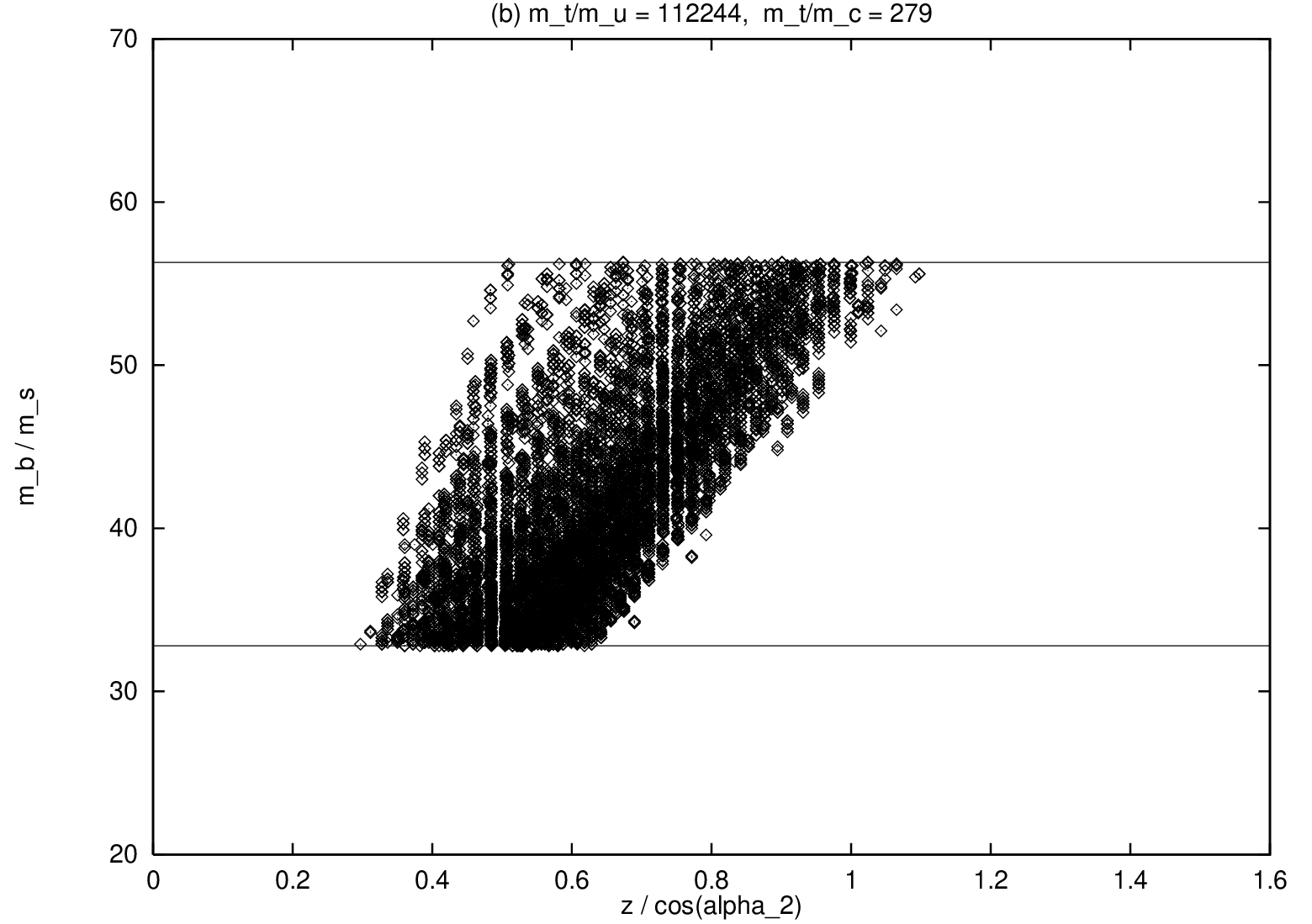}}
\caption{The mass ratio $m_b/m_s$ as a function of $z/\cos\alpha_2$.
 The allowed region is between two horizontal lines 
 ($32.8 < m_b/m_s < 56.3$).
 We have set $\hat{m}_u^{-1}$ to be $112244$ and $\hat{m}_c^{-1}$ to be
 $(a) 394$ (central value) $(b) 279$ (lowest value).}
\label{fig-m_b-m_s}
\end{figure}

\begin{figure}
\epsfxsize=10cm
\centerline{\epsfbox{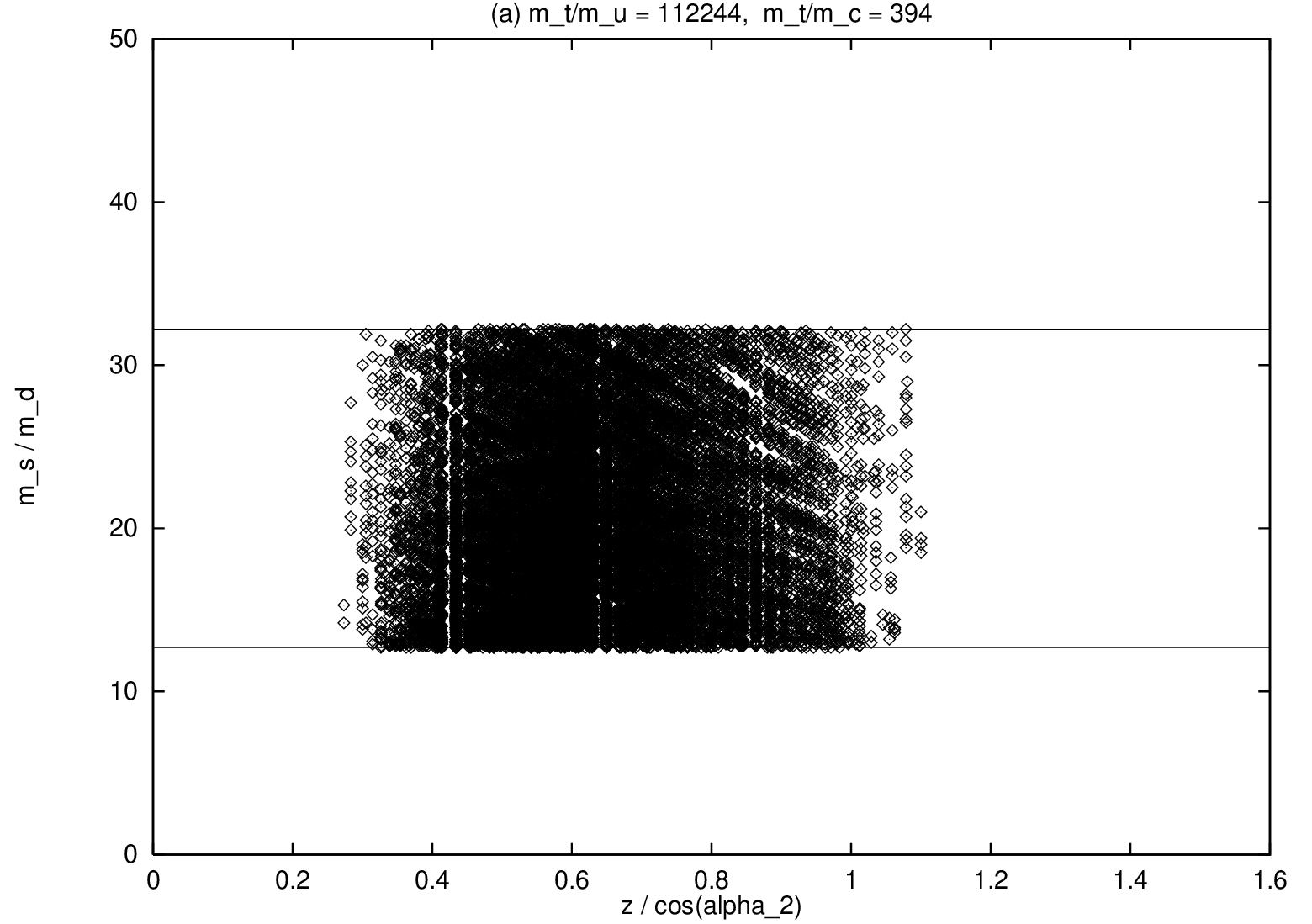}}
\vspace{1cm}
\epsfxsize=10cm
\centerline{\epsfbox{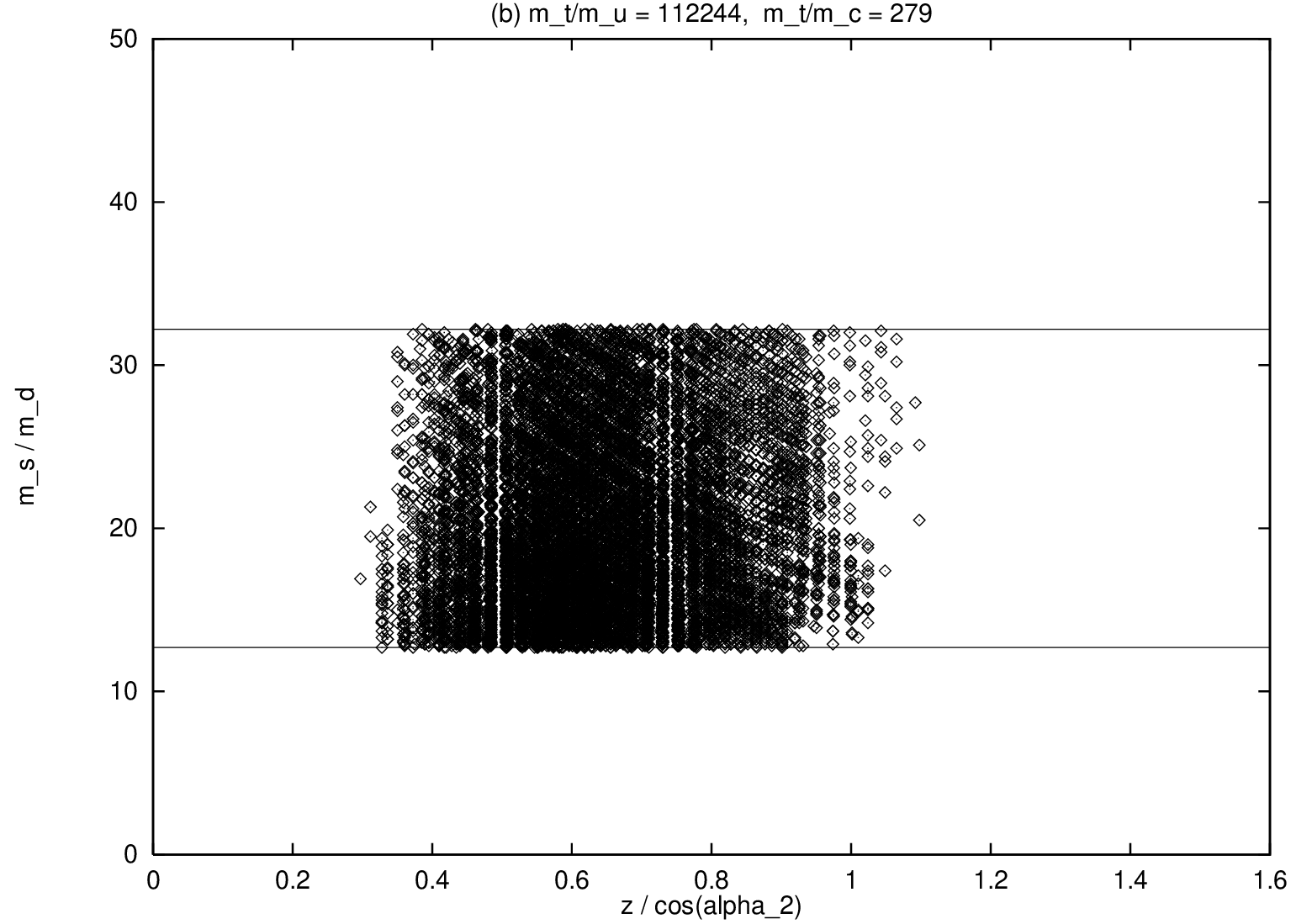}}
\caption{The mass ratio $m_s/m_d$ as a function of $z/\cos\alpha_2$.
 The allowed region is between two horizontal lines 
 ($12.7 < m_b/m_s < 32.2$).
 We have set $\hat{m}_u^{-1}$ to be $112244$ and $\hat{m}_c^{-1}$ to be
 $(a) 394$ (central value) $(b) 279$ (lowest value).}
\label{fig-m_s-m_d}
\end{figure}

Finally, we plot the down-type quark mass ratios $m_b/m_s$ and $m_s/m_d$
as functions of $z/\cos\alpha_2$ in Fig.~\ref{fig-m_b-m_s} and
\ref{fig-m_s-m_d}, respectively.
We have ascertained that $m_b/m_s$ and $m_s/m_d$ take any value in a
range $32.8 < m_b/m_s < 56.3$ and $12.7 < m_s/m_d < 32.2$, so that there 
is no relation between them.
Note that the allowed region of $z/\cos\alpha_2$ is hardly dependent on
$\hat{m}_c^{-1}$, since $z$ and $\cos\alpha_2$ respond to the change of
$\hat{m}_c^{-1}$ in the same way.
Thus, our qualitative discussion below is almost independent of  
$\hat{m}_c^{-1}(M_{\rm GUT})$.
The quantity $z/\cos\alpha_2$ almost corresponds to the tangent of the 
MNS mixing angle, $\tan\theta_{\mu 3}$, between second and third
generations, since it dictates the mixing between left-handed charged
leptons of second and third generations.
Actual mixing angle is the sum of it and an additional contribution from
neutrino mass matrix $M_{\nu}$.
Considering that $\delta_z$ is given by Eq.~(\ref{neutrino_delta}) and
$h_3 = O(1)$, the additional contribution can be of order the Cabibbo
angle.
Thus, if two contributions are added up constructively, the region 
$z/\cos\alpha_2 \gsim 0.4$ is consistent with the observed near maximal
mixing of the atmospheric neutrino oscillation 
($\nu_{\mu} \leftrightarrow \nu_{\tau}$) 
\cite{atm-osc, SuperK}.\footnote{
Even if we have no contribution from neutrino sector, the region 
$z/\cos\alpha_2 \gsim 0.6$ is consistent with the observed near maximal
mixing of the atmospheric neutrino oscillation, 
$\sin^2 2\theta_{\mu 3} \gsim 0.8$.
In this case, relatively large value of $m_b / m_s$ is favorable.}
We find that significant region is consistent with the large angle
between $\nu_{\mu}$ and $\nu_{\tau}$.

\section{The masses and mixings in the lepton sector}

In this section, we discuss the lepton sector of the model taking
$\hat{m}_u^{-1}$ and $\hat{m}_c^{-1}$ to be their central values for
simplicity.
First, we begin with the charged-lepton masses whose precise values are
known experimentally as in Table~\ref{quark-lepton_mass}.
The mass matrix for the charged leptons is different from that for the
down-type quarks due to the presence of nonrenormalizable interactions
Eq.~(\ref{GJ-superpot}).
We assume that the nonrenormalizable interactions
Eq.~(\ref{GJ-superpot}) modify only $(1,2)$ and $(2,2)$ components of
the mass matrices.
Thus, we search a parameter region of $x_l$ and $y_l$ which can well
reproduce the known lepton mass ratios at the GUT scale, 
$m_{\tau}/m_{\mu} = 17.07$ and $m_{\mu}/m_e = 211.1$, using the values
of $z$, $\alpha_1$ and $\alpha_2$ obtained in section 3 which reproduce 
the quark mass ratios and the CKM parameters.

\begin{figure}
\epsfxsize=10cm
\centerline{\epsfbox{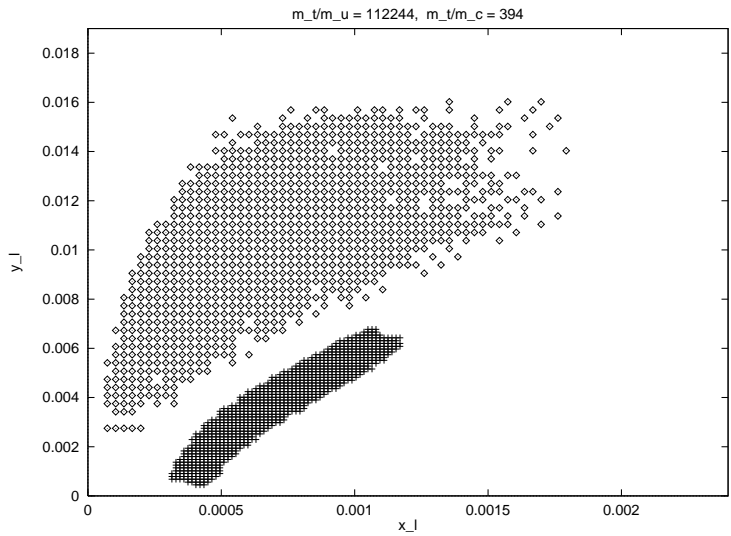}}
\caption{The parameter region of $x_l$ and $y_l$ where all lepton mass
 ratios are reproduced.
 The hatched region represents that of $x_d$ and $y_d$ given in
 Fig.~\ref{fig-x_d-y_d} $(a)$.
 We have set $\hat{m}_u^{-1} = 112244$ and $\hat{m}_c^{-1} = 394$.}
\label{fig-x_l-y_l}
\end{figure}

The resulting region of $x_l$ and $y_l$ is given in
Fig.~\ref{fig-x_l-y_l} together with that of $x_d$ and $y_d$ obtained in
Fig.~\ref{fig-x_d-y_d}.
We find that there is significant region consistent with the observed
charged-lepton masses.
The preferred region is $x_l \sim x_d$ and $y_l \sim 3y_d$, which agrees
with the earlier observation \cite{G-J}.
Thus, we conclude that the presence of nonrenormalizable interactions
Eq.~(\ref{GJ-superpot}) with $g'_i = O(1)$ is sufficient to remove
unwanted SU(5)$_{\rm GUT}$ mass relations, 
$m_{\mu} = m_s$ and $m_e = m_d$, also in the present model.

Next, we discuss the neutrino masses and mixings qualitatively.
We call the mass eigenstates for three neutrinos $\nu_1$, $\nu_2$ and
$\nu_3$ such that $m_{\nu_1} < m_{\nu_2} < m_{\nu_3}$.
In the present model, we have a mass hierarchy 
$m_{\nu_2} \ll m_{\nu_3}$ due to the fact that 
$\delta_x, \delta_y, \delta_z \lsim 10^{-2}$.
Then, the data of atmospheric neutrino oscillation 
($\nu_{\mu} \leftrightarrow \nu_{\tau}$) from Super-Kamiokande implies
that $m_{\nu_3} \simeq 7 \times 10^{-2}~\EV$.
The neutrino mass matrix $M_{\nu}$ is given by 
$M_{\nu} = M_{\nu D}^{\rm T} M_N^{-1} M_{\nu D}$.
Thus, the scale $M_R$ of the Majorana masses is determined as 
$M_R \simeq 10^{12}-10^{13}~\GEV$, which is close to the natural scale
derived from Eq.~(\ref{Majorana_mass}) with $j_{33} = O(1)$.
The observed maximal mixing is also naturally obtained as a result of
fitting quark masses and mixings.
In the following, we discuss the implications of the model on the solar
neutrino deficit.
For simplicity, we ignore CP-violating effects in the lepton sector and
take all values appear in the neutrino mass matrix to be real,
$\varphi_x = \varphi_y = \varphi_z = \varphi_t = \varphi_u = \varphi_v 
= \varphi_w = 0$ or $\pi$.

The observed solar neutrino deficit is explained by either 
matter-enhanced neutrino oscillation (MSW \cite{MSW}) solution or the
vacuum oscillation solution \cite{VO} 
($\nu_e \leftrightarrow \nu_{\mu}, \nu_{\tau}$).
Then, the allowed regions of mass squared differences and mixing angles
are as shown in Table~\ref{neutrino_angle} \cite{sol-osc, sol-osc-VO}.
The MSW solution has two distinct regions, that is, the small and the
large angle ones.
\begin{table}
\begin{center}
\begin{tabular}{|c|c|c|c|}  \hline 
          solutions &     $\delta m_{\rm sol}^2$    & 
          $\sin^2 2\theta_{\rm sol}$ & $m_{\nu_2}/m_{\nu_3}$ \\ \hline
 small angle MSW    & $(4-12) \times 10^{-6}~\EV^2$ & 
          $(2-12) \times 10^{-3}$    & $\sim 0.03$     \\ \hline
 large angle MSW    & $(8-25) \times 10^{-6}~\EV^2$ & 
          $0.5 - 0.8$                & $\sim 0.05$     \\ \hline
                    & $(6-7) \times 10^{-10}~\EV^2$ & 
          $0.8 - 1.0$                & $\sim 10^{-4}$  \\
 vacuum oscillation & $(4-5) \times 10^{-10}~\EV^2$ & 
          $0.7 - 1.0$                & $\sim 10^{-4}$  \\
                    & $(5-9) \times 10^{-11}~\EV^2$ & 
          $0.6 - 0.9$                & $\sim 10^{-4}$  \\ \hline
\end{tabular}
\end{center}
\caption{The allowed regions of mass squared differences and 
 mixing angles which reproduce the observed solar neutrino deficit 
 in terms of the neutrino oscillation \cite{sol-osc, sol-osc-VO}.
 We have also shown mass ratios $m_{\nu_2}/m_{\nu_3}$ under the 
 mass hierarchy $m_{\nu_2} \ll m_{\nu_3}$.}
\label{neutrino_angle}
\end{table}
Below, we investigate which solutions the model can realize in two cases
that the neutrino Majorana mass matrix $M_N$ is proportional to the unit
matrix or that it has a hierarchy consistent with the observation that
$\psi_i({\bf 16})$ have a source of hierarchies.
The RG effects between the GUT scale and the weak scale are negligible
for our qualitative argument, so that we evaluate the masses and mixing
angles at the GUT scale and compare them with the observed values given
in Table~\ref{neutrino_angle}.

\subsection{The case of unit Majorana mass matrix}

We first consider the unit Majorana mass matrix case;
\begin{eqnarray}
  M_N = M_R \pmatrix{
          1 & 0 & 0  \cr
          0 & 1 & 0  \cr
          0 & 0 & 1  \cr}.
\end{eqnarray}
Here, we simply assume that $M_N$ is proportional to the unit matrix for 
some reasons in the basis that the original Yukawa couplings are
diagonal, and regard $\delta_x$, $\delta_y$ and $\delta_z$ as free
parameters.
Then, the neutrino mass matrix, 
$M_{\nu} = M_{\nu D}^{\rm T} M_N^{-1} M_{\nu D}$, is given as
\begin{eqnarray}
  M_{\nu} &=& \frac{m_t^2}{M_R}
  \pmatrix{
          -\hat{m}_u s_1 & \hat{m}_c c_1 & 0  \cr
          \delta_x & \delta_y & \delta_z  \cr
          \hat{m}_u c_1 s_2 & 
              \hat{m}_c s_1 s_2 & 
              c_2  \cr}
  \pmatrix{
          -\hat{m}_u s_1 & \delta_x &
              \hat{m}_u c_1 s_2  \cr
          \hat{m}_c c_1 & \delta_y &
              \hat{m}_c s_1 s_2  \cr
          0 & \delta_z & c_2  \cr} \nonumber\\
  &\simeq& \frac{m_t^2}{M_R}
  \pmatrix{
          \hat{m}_u^2 c_1^2 & 
              -\hat{m}_u\delta_x + \hat{m}_c\delta_y c_1 &
              \hat{m}_c^2 c_1  \cr
          -\hat{m}_u\delta_x + \hat{m}_c\delta_y c_1 &
              \delta_x^2 + \delta_y^2 + \delta_z^2 &
              \hat{m}_u\delta_x c_1 + \hat{m}_c\delta_y 
              + \delta_z c_2  \cr
          \hat{m}_c^2 c_1 &
              \hat{m}_u\delta_x c_1 + \hat{m}_c\delta_y 
              + \delta_z c_2 &
              c_2^2  \cr},
\label{nu_unit}
\end{eqnarray}
where $c_i \equiv \cos\alpha_i$ and $s_i \equiv \sin\alpha_i$ ($i=1,2$).

The contribution to the mixing angle between second and third
generations from the neutrino mass matrix Eq.~(\ref{nu_unit}) is given
by $\theta^{\nu}_{\mu 3} \sim \delta_z / c_2$.
In view of $c_2 = O(0.1)$, we choose $\delta_z \sim \epsilon^2$ to give
naturally the near maximal mixing in the atmospheric neutrino
oscillation (see discussion at the end of section 3).
Then, the relevant mass ratio and mixing to the solar neutrino
oscillation are given as
\begin{eqnarray}
  && \frac{m_{\nu_2}}{m_{\nu_3}} \sim \frac{\delta_y^2}{c_2^2}
         \sim \frac{\delta_y^2}{\epsilon^2}, 
\label{ratio_unit}\\
  && \theta^{\nu}_{e 2} \sim \frac{\hat{m}_c c_1}{\delta_y}
         \sim \frac{\epsilon^3}{\delta_y},
\label{angle_unit}
\end{eqnarray}
as long as $\delta_x \lsim \delta_y$ and $\delta_y$ is not very small
compared with $\delta_z$.

From Eq.~(\ref{ratio_unit}), we find that if we choose 
$\delta_y \sim \epsilon^2$ we obtain the mass ratio consistent with the
MSW solutions.
Then, the mixing angle $\theta^{\nu}_{e 2}$ is of order of $\epsilon$
from Eq.~(\ref{angle_unit}).
The actual mixing angle is the sum of $\theta^{\nu}_{e 2}$ and the
corresponding contribution from charged-lepton mass matrix
$\theta^{l}_{e 2}$.
Since analyses of the previous section indicate 
$\theta^{l}_{e 2} \simeq 0.04-0.19$, the resulting mixing angle is
consistent with the small angle MSW solution if two contributions are
added up destructively, while it cannot reach that of the large angle
MSW solution even if two contributions are added up constructively.
The required cancellation is rather mild such that the reduction of
factor two or three is sufficient.
We also mention that the mixing angle $\theta_{e 3}$ is small, so that
it is not conflict with the result of long baseline reactor neutrino
oscillation experiment (CHOOZ), $\theta_{e 3} \lsim 0.22$ \cite{CHOOZ}.

If we choose $\delta_y \sim \epsilon^3$, on the other hand, both mass
ratio and mixing angle are consistent with vacuum oscillation solution.
The reason for this coincidence can be traced back to the fact that the
required neutrino mass ratio $m_{\nu_2}/m_{\nu_3} \sim 10^{-4}$ is
approximately equal to $(m_c/m_t)^2$ as predicted by a simple (unit
$M_N$) see-saw model in the SO(10) GUT.
That is, if we intend to obtain large solar-neutrino mixing angle we
have to choose $\delta_y \simeq \hat{m}_c c_1$, and then we necessarily
have the mass ratio 
$m_{\nu_2}/m_{\nu_3} \sim \delta_y^2/c_2^2 \sim \hat{m}_c^2$.
The resulting MNS matrix has the so-called bi-maximal form 
\cite{bi-maximal, previous} and evades the constraint from CHOOZ
experiment.
We note that this vacuum oscillation solution is realized with 
$k_1 : k_2 : k_3 \simeq \epsilon^2 : \epsilon : 1$ which is implied by
the hierarchy Eqs.~(\ref{hierarchy_1}, \ref{hierarchy_2}) 
(see Eq.~(\ref{nonren_nuD})).

\subsection{The case of hierarchical Majorana mass matrix}

If $\psi_i({\bf 16})$ have a source of hierarchies indicated by 
Eqs.~(\ref{hierarchy_1}, \ref{hierarchy_2}), it is natural to think 
that the Majorana mass matrix $M_N$ also has hierarchy consistent with
it.
Thus, we consider the case where the Majorana mass matrix has the
hierarchical form;
\begin{eqnarray}
  M_N \simeq M_R \pmatrix{
          \epsilon^4 & \epsilon^3 & \epsilon^2  \cr
          \epsilon^3 & \epsilon^2 & \epsilon    \cr
          \epsilon^2 & \epsilon   & 1           \cr},
\end{eqnarray}
up to order-one coefficients.

In this subsection, we stick to the possibility that $\psi_i({\bf 16})$
have a source of hierarchies indicated by 
Eqs.~(\ref{hierarchy_1}, \ref{hierarchy_2}), so that we assume that 
$k_1 : k_2 : k_3 \simeq \epsilon^2 : \epsilon : 1$.
Considering that $\delta_z \sim \epsilon^2$ in order to give naturally
the near maximal mixing between $\nu_{\mu}$ and $\nu_{\tau}$, elements
of the $M_{\nu D}$ have the following order of magnitude:
\begin{eqnarray}
  \delta_x \simeq \epsilon^4, \qquad 
  \delta_y \simeq \epsilon^3, \qquad
  \delta_z \simeq \epsilon^2.
\end{eqnarray}
Then, the neutrino mass matrix, 
$M_{\nu} = M_{\nu D}^{\rm T} M_N^{-1} M_{\nu D}$, is given as
\begin{eqnarray}
  M_{\nu} &\simeq& \frac{m_t^2}{M_R}
  \pmatrix{
          \epsilon^{-4}\tilde{m}_u^2 & 
              \tilde{m}_u &
              \epsilon^{-3}\tilde{m}_u\tilde{m}_c  \cr
          \tilde{m}_u &
              \epsilon^4 &
              \epsilon\tilde{m}_c  \cr
          \epsilon^{-3}\tilde{m}_u\tilde{m}_c &
              \epsilon\tilde{m}_c &
              \epsilon^{-2}\tilde{m}_c^2  \cr},
\end{eqnarray}
where $\tilde{m}_u \equiv \hat{m}_u + \epsilon\hat{m}_c c_1$ and 
$\tilde{m}_c \equiv \hat{m}_c + \epsilon c_2$.
This gives the relevant mass ratio and mixing to the solar neutrino
oscillation as
\begin{eqnarray}
  && \frac{m_{\nu_2}}{m_{\nu_3}} \sim \frac{\epsilon^6}{\tilde{m}_c^2}
         = \frac{\epsilon^6}{(\hat{m}_c + \epsilon c_2)^2}, 
\label{ratio_hier}\\
  && \theta^{\nu}_{e 2} \sim \frac{\tilde{m}_u}{\epsilon^4}
         = \frac{\hat{m}_u + \epsilon\hat{m}_c c_1}{\epsilon^4}.
\label{angle_hier}
\end{eqnarray}

From Eq.~(\ref{ratio_hier}), we obtain 
$m_{\nu_2}/m_{\nu_3} \simeq 10^{-2}-10^{-1}$, which is consistent with
both small and large angle MSW solutions.
Also, the angle is given as $\theta^{\nu}_{e 2} \simeq 0.1-1$ from
Eq.~(\ref{angle_hier}), so that we can reproduce the large angle MSW
solution naturally.
To reproduce the small angle solution, $\theta^{\nu}_{e 2}$ and 
$\theta^{l}_{e 2}$ have to be added up destructively, but the required
cancellation is mild.
(The reduction of factor three to five is sufficient.)
Thus, we conclude that the the model can accommodate both large and
small angle MSW solutions in the case of the hierarchical Majorana mass
matrix.

\section{Conclusions}

In this paper, we have investigated the SO(10) GUT model with generation 
flipping.
The model contains one extra matter multiplet $\psi({\bf 10})$ in
addition to the usual matter multiplets $\psi_i({\bf 16})$, and it mixes 
with $\psi_i({\bf 16})$ when the SO(10)$_{\rm GUT}$ is broken down to
SU(5) by the vacuum expectation values of $\chi({\bf 16})$ and
$\bar{\chi}({\bf 16}^*)$ fields.\footnote{
This structure may be realized in a framework of E$_6$ GUT \cite{E6}.}
The low-energy quarks and leptons are three linear combinations of
$\psi_i({\bf 16})$ and $\psi({\bf 10})$.

We have found the parameter region of the model in which the observed
quark masses and mixings are well reproduced.
The required hierarchies of the coupling constants are consistent with
the observation that only $\psi_i({\bf 16})$ have a source of
hierarchies such as U(1) flavor symmetry charges.
As for the lepton sector, the model can reproduce the charged-lepton
masses if there are suitable nonrenormalizable interactions which 
introduce SU(5) breaking effects into $M_{d/l}$.
The obtained charged-lepton mass matrix indicates that the mixing
between second and third generations tends to be large in the lepton
sector, which is consistent with the observed maximal mixing of the
atmospheric neutrino oscillation.

We have also discussed neutrino masses and mixings qualitatively.
We have considered two cases that the Majorana mass matrix $M_N$ for the 
right-handed neutrinos is proportional to the unit matrix and it has the 
hierarchical form.
In the former case, the model can accommodate small angle MSW solution
and vacuum oscillation solution to the solar neutrino deficit, depending 
on the values of the coupling constants $k_i$.
The vacuum oscillation solution is realized when $k_i$ have the
hierarchy indicated in the quark sector, and the resulting MNS mixing
matrix has the form of so-called bi-maximal mixing.

In the latter case, the model can accommodate both small and large angle
MSW solutions and the parameter region is consistent with the hierarchy
required in the quark sector.
It is remarkable that the single anomalous U(1)$_X$ flavor symmetry
with the $\phi$ field which has the vacuum expectation value of order
$10^{17}~\GEV$ realizes this possibility.\footnote{
The nonrenormalizable operators in Eq.~(\ref{GJ-superpot}) might be
generated via exchanges of some heavy fields, so that they are
suppressed by the mass scale of the heavy fields instead of $M_G$.
We think that the masses for these heavy fields, the GUT scale $V$ and
the mass $M$ are not related to the anomalous U(1)$_X$ breaking scale,
$\vev{\phi}$.}
Then, the operators in the superpotential have suppression factors of
appropriate powers of $\vev{\phi}/M_G \simeq \epsilon$ depending on
their U(1)$_X$ charges \cite{anom-U1}.
An example of the U(1)$_X$ charges for various fields is given in
Table~\ref{U1X_charge}.\footnote{
Under the charge assignment of Table~\ref{U1X_charge}, the mass $M$ in
Eq.~(\ref{Higgs_pot}) should be $10^{17}~\GEV$.
On the other hand, if $\psi({\bf 10})$ has the U(1)$_X$ charge $1$, $M$
should be the GUT scale, $10^{16}~\GEV$.}
\begin{table}
\begin{center}
\begin{tabular}{|c|ccccccc|c|}  \hline 
  & $\psi_1({\bf 16})$       & $\psi_2({\bf 16})$ & $\psi_3({\bf 16})$ & 
    $\psi({\bf 10})$         & $H({\bf 10})$      & $\chi({\bf 16})$   & 
    $\bar{\chi}({\bf 16}^*)$ & $\phi$             \\ \hline
  U(1)$_X$ & 2 & 1 & 0 & 0 & 0 & 0 & 0 & $-1$ \\ \hline
\end{tabular}
\end{center}
\caption{U(1)$_X$ charges.
 The other fields have vanishing charges.}
\label{U1X_charge}
\end{table}
Note, however, that this U(1)$_X$ alone is incomplete to obtain the
superpotential of the present model, since we have to suppress unwanted
terms such as $M_G\, \psi({\bf 10})\, \psi({\bf 10})$ by hand.
Thus, additional symmetries are required to make the model natural.

To summarize, we have found that in the present model the observed solar
neutrino deficit is explained by different neutrino-oscillation
solutions depending on the form of the Majorana mass matrix $M_N$.
We hope that future experiments will distinguish among different
solutions and we will be able to approach the structure of the Majorana
mass matrix for the right-handed neutrinos.

\section*{Acknowledgments}

We would like to thank T.~Yanagida for valuable discussions and comments 
on the present work.
We also thank J.~Hashiba, K.~Kurosawa, N.~Okamura and J.~Sato for
discussions. 
Y.N. is supported by the Japan Society for the Promotion of Science.

\appendix
\section{The mass matrices in the model}

In this appendix, we give a detailed derivation of the mass matrices in
the model, clarifying a phase convention we have used.

The model contains matter fields $\psi_i({\bf 16}), \psi({\bf 10})$
and Higgs fields $H({\bf 10}), H({\bf 16})$.
After the spontaneous breakdown of SO(10)$_{\rm GUT}$ to SU(5), these
fields decompose as
\begin{eqnarray}
  \left\{ \matrix{
  {\bf 10}_1 &+& {{\bf 5}^*_1}' &+& {\bf 1}_1 &=& \psi_1({\bf 16}), \cr
  {\bf 10}_2 &+& {{\bf 5}^*_2}' &+& {\bf 1}_2 &=& \psi_2({\bf 16}), \cr
  {\bf 10}_3 &+& {{\bf 5}^*_3}' &+& {\bf 1}_3 &=& \psi_3({\bf 16}), \cr
  {\bf 5}_{\psi} &+& {\bf 5}^*_{\psi} &&      &=& \psi({\bf 10}), \cr}
  \right.
\end{eqnarray}
\begin{eqnarray}
  \left\{ \matrix{
  {\bf 5}_H &+& {\bf 5}^*_{H({\bf 10})} && &=& H({\bf 10}), \cr
  {\bf 10}_{H({\bf 16})} &+& {\bf 5}^*_{H({\bf 16})} &+& 
        {\bf 1}_{H({\bf 16})} &=& H({\bf 16}). \cr}
  \right.
\end{eqnarray}

The original Yukawa couplings are given by the superpotential
Eq.~(\ref{diagonal_Yukawa}) and the Majorana masses for the right-handed
neutrinos by Eq.~(\ref{Majorana_mass}).
At this stage, the quark and lepton masses are written as
\begin{eqnarray}
  W &=& \frac{1}{2} \pmatrix{
          {\bf 10}_1\, {\bf 10}_2\, {\bf 10}_3  \cr}
        \pmatrix{
          h_1\vev{{\bf 5}_H} & 0 & 0  \cr
          0 & h_2\vev{{\bf 5}_H} & 0  \cr
          0 & 0 & h_3\vev{{\bf 5}_H}  \cr}
        \pmatrix{
          {\bf 10}_1 \cr  {\bf 10}_2 \cr  {\bf 10}_3  \cr} 
  \nonumber\\
    && + \pmatrix{
          {\bf 10}_1\, {\bf 10}_2\, {\bf 10}_3  \cr}
        \pmatrix{
          h_1\vev{{\bf 5}^*_{H({\bf 10})}} & 0 & 0  \cr
          0 & h_2\vev{{\bf 5}^*_{H({\bf 10})}} & 0  \cr
          0 & 0 & h_3\vev{{\bf 5}^*_{H({\bf 10})}}  \cr}
        \pmatrix{
          {{\bf 5}^*_1}' \cr  {{\bf 5}^*_2}' \cr  {{\bf 5}^*_3}'  \cr}
  \nonumber\\
    && + \pmatrix{
          {\bf 1}_1\, {\bf 1}_2\, {\bf 1}_3  \cr}
        \pmatrix{
          h_1\vev{{\bf 5}_H} & 0 & 0  \cr
          0 & h_2\vev{{\bf 5}_H} & 0  \cr
          0 & 0 & h_3\vev{{\bf 5}_H}  \cr}
        \pmatrix{
          {{\bf 5}^*_1}' \cr  {{\bf 5}^*_2}' \cr  {{\bf 5}^*_3}'  \cr}
  \nonumber\\
    && + \frac{1}{2} \pmatrix{
          {\bf 1}_1\, {\bf 1}_2\, {\bf 1}_3  \cr}
        \pmatrix{
          M_R\,\tilde{\jmath}_{11} & M_R\,\tilde{\jmath}_{12} & 
               M_R\,\tilde{\jmath}_{13}  \cr
          M_R\,\tilde{\jmath}_{21} & M_R\,\tilde{\jmath}_{22} & 
               M_R\,\tilde{\jmath}_{23}  \cr
          M_R\,\tilde{\jmath}_{31} & M_R\,\tilde{\jmath}_{32} & 
               M_R\,\tilde{\jmath}_{33}  \cr}
        \pmatrix{
          {\bf 1}_1 \cr  {\bf 1}_2 \cr  {\bf 1}_3  \cr}, 
\label{mass_1}
\end{eqnarray}
in terms of SU(5) language.
Here, $\tilde{\jmath}_{ij} = j_{ij}\,/|j_{33}|$ and $M_R$ is given as 
$M_R = |j_{33}|\, V^2 / M_G$.
If we assume that the Majorana mass matrix is proportional to the unit
matrix, $j_{ij} = j\, \delta_{ij}$, in the basis that the
original Yukawa couplings are diagonal, $\tilde{\jmath}_{ij}$ and $M_R$
are reduced to $\tilde{\jmath}_{ij} = \delta_{ij}$ and 
$M_R = j\, V^2 / M_G$.

The superpotential Eq.~(\ref{mass_matter}) mixes $\psi({\bf 10})$ with 
$\psi_i({\bf 16})$.
This term can be written in terms of SU(5) language as 
\begin{eqnarray}
  W = \sqrt{|f_1|^2+|f_2|^2+|f_3|^2} \vev{\chi({\bf 16})}
      \tilde{\bf 5}^*\, {\bf 5}_{\psi},
\label{heavy_matter}
\end{eqnarray}
where $\tilde{\bf 5}^*$ is a linear combination of ${{\bf 5}^*_i}'$,
$\tilde{\bf 5}^* = \sum_{i=1}^{3} f_i \e^{i\eta_i}\, 
{{\bf 5}^*_i}' / \sqrt{|f_1|^2+|f_2|^2+|f_3|^2}$.
We call remaining two linear combinations ${\bf 5}^*_1$ and 
${\bf 5}^*_3$.
The fields $\tilde{\bf 5}^*, {\bf 5}^*_1$ and ${\bf 5}^*_3$
are related to ${{\bf 5}^*_i}'$ by a unitary transformation.
Thus, using arbitrariness of defining ${\bf 5}^*_1$ and ${\bf 5}^*_3$
and phases of the fields $\tilde{\bf 5}^*, {\bf 5}^*_1$ and 
${\bf 5}^*_3$, we can write this relation as
\begin{eqnarray}
  \pmatrix{{{\bf 5}^*_1}' \cr
           {{\bf 5}^*_2}' \cr
           {{\bf 5}^*_3}' \cr}
  = 
  \pmatrix{
          \e^{-i\delta_1-i\delta_2} & 0 &        0        \cr
          0 & \e^{-i\delta_1+i\delta_2} &        0        \cr
                0       &       0       & \e^{2i\delta_1} \cr}
  \pmatrix{
          \cos\alpha_1 & -\sin\alpha_1 & 0 \cr
          \sin\alpha_1 &  \cos\alpha_1 & 0 \cr
                0      &        0      & 1 \cr} \nonumber\\
  \times
  \pmatrix{
                   \cos\alpha_2       & 0 & \e^{-i\delta_3}\sin\alpha_2 \cr
                        0             & 1 &               0             \cr
          -\e^{i\delta_3}\sin\alpha_2 & 0 &          \cos\alpha_2       \cr}
  \pmatrix{\tilde{\bf 5}^* \cr
           {\bf 5}^*_1     \cr
           {\bf 5}^*_3     \cr},
\label{unitary_rel}
\end{eqnarray}
without a loss of generality.
Here, $\alpha_1, \alpha_2$ and $\delta_i$ are functions of $f_i$ and
$\eta_i$.
Note that $\tilde{\bf 5}^*$ receives a GUT scale mass together with
${\bf 5}_{\psi}$ by Eq.~(\ref{heavy_matter}), so that the low-energy
quark and lepton fields belonging to ${\bf 5}^*$ of SU(5) are 
${\bf 5}^*_1$, ${\bf 5}^*_3$ and ${\bf 5}^*_{\psi}$.

The $\psi({\bf 10})$ is coupled with Higgs and $\psi_i({\bf 16})$ by
superpotentials Eqs.~(\ref{down-type_giving}, \ref{nonren_nuD}).
Then, from Eqs.~(\ref{mass_1}, \ref{unitary_rel}) we obtain
down-type quark (charged-lepton) mass matrix $M_{{\bf 10}-{\bf 5}^*}$
and neutrino Dirac mass matrix $M_{{\bf 1}-{\bf 5}^*}$ defined by
\begin{eqnarray}
  W = \pmatrix{
          {\bf 10}_1\, {\bf 10}_2\, {\bf 10}_3  \cr} M_{{\bf 10}-{\bf 5}^*}
      \pmatrix{
          {\bf 5}^*_1 \cr  {\bf 5}^*_{\psi} \cr  {\bf 5}^*_3  \cr}
    + \pmatrix{
          {\bf 1}_1\, {\bf 1}_2\, {\bf 1}_3  \cr} M_{{\bf 1}-{\bf 5}^*}
      \pmatrix{
          {\bf 5}^*_1 \cr  {\bf 5}^*_{\psi} \cr  {\bf 5}^*_3  \cr},
\end{eqnarray}
as
\begin{eqnarray}
  && M_{{\bf 10}-{\bf 5}^*} = \nonumber\\
  && \pmatrix{
     -h_1\vev{{\bf 5}^*_{H({\bf 10})}}\e^{-i\delta_1-i\delta_2}\sin\alpha_1 &
      g_1\vev{{\bf 5}^*_{H({\bf 16})}}\e^{i\phi_1} &
      h_1\vev{{\bf 5}^*_{H({\bf 10})}}\e^{-i\delta_1-i\delta_2-i\delta_3}
          \cos\alpha_1 \sin\alpha_2  \cr
      h_2\vev{{\bf 5}^*_{H({\bf 10})}}\e^{-i\delta_1+i\delta_2}\cos\alpha_1 &
      g_2\vev{{\bf 5}^*_{H({\bf 16})}}\e^{i\phi_2} &
      h_2\vev{{\bf 5}^*_{H({\bf 10})}}\e^{-i\delta_1+i\delta_2-i\delta_3}
          \sin\alpha_1 \sin\alpha_2  \cr
      0 &
      g_3\vev{{\bf 5}^*_{H({\bf 16})}}\e^{i\phi_3} &
      h_3\vev{{\bf 5}^*_{H({\bf 10})}}\e^{2i\delta_1}\cos\alpha_2  \cr},
  \nonumber\\
  && M_{{\bf 1}-{\bf 5}^*} = \nonumber\\
  && \pmatrix{
     -h_1\vev{{\bf 5}_H}\e^{-i\delta_1-i\delta_2}\sin\alpha_1 &
      k_1\vev{{\bf 5}_H}\frac{V}{M_G}\e^{i\varphi_1} &
      h_1\vev{{\bf 5}_H}\e^{-i\delta_1-i\delta_2-i\delta_3}
          \cos\alpha_1 \sin\alpha_2  \cr
      h_2\vev{{\bf 5}_H}\e^{-i\delta_1+i\delta_2}\cos\alpha_1 &
      k_2\vev{{\bf 5}_H}\frac{V}{M_G}\e^{i\varphi_2} &
      h_2\vev{{\bf 5}_H}\e^{-i\delta_1+i\delta_2-i\delta_3}
          \sin\alpha_1 \sin\alpha_2  \cr
      0 &
      k_3\vev{{\bf 5}_H}\frac{V}{M_G}\e^{i\varphi_3} &
      h_3\vev{{\bf 5}_H}\e^{2i\delta_1}\cos\alpha_2  \cr}.
\label{mass_2}
\end{eqnarray}

SU(5)-${\bf 5}^*$ of the Higgs fields $H({\bf 10})$ and $H({\bf 16})$
are also mixed by the superpotential Eq.~(\ref{Higgs_pot}).
One linear combination $\tilde{\bf 5}^*_H$ gets a GUT scale mass
together with SU(5)-${\bf 5}$ of $\bar{H}({\bf 16}^*)$ and the other
combination ${\bf 5}^*_H$ remains as a massless Higgs in the standard
SU(5)$_{\rm GUT}$.
We can parametrize these fields as
\begin{eqnarray}
  \pmatrix{{\bf 5}^*_{H({\bf 10})}  \cr
           {\bf 5}^*_{H({\bf 16})}  \cr}
  = 
  \pmatrix{
          \cos\theta & -\sin\theta  \cr
          \sin\theta &  \cos\theta  \cr}
  \pmatrix{{\bf 5}^*_H  \cr
           \tilde{\bf 5}^*_H  \cr},
\end{eqnarray}
where $\tan\theta = -\lambda\, V/M$.
That is,
\begin{eqnarray}
  \vev{{\bf 5}^*_{H({\bf 10})}} &=& \cos\theta \vev{{\bf 5}^*_H}, \nonumber\\
  \vev{{\bf 5}^*_{H({\bf 16})}} &=& \sin\theta \vev{{\bf 5}^*_H}.
\label{Higgs_VEV}
\end{eqnarray}
Substituting Eq.~(\ref{Higgs_VEV}) for Eq.~(\ref{mass_2}) and combining
with Eq.~(\ref{mass_1}), we obtain complete mass matrices.
Defining
\begin{eqnarray}
  m_u \equiv h_1\, \vev{{\bf 5}_H}, \qquad
  m_c \equiv h_2\, \vev{{\bf 5}_H}, \qquad
  m_t \equiv h_3\, \vev{{\bf 5}_H}, \qquad
  \tan\beta \equiv \vev{{\bf 5}_H} / \vev{{\bf 5}^*_H},
\end{eqnarray}
the quark and lepton masses are written as
\begin{eqnarray}
  W &=& \pmatrix{
          u\, c\, t  \cr}
        \pmatrix{
          m_u & 0 & 0  \cr
          0 & m_c & 0  \cr
          0 & 0 & m_t  \cr}
        \pmatrix{
          \bar{u} \cr  \bar{c} \cr  \bar{t}  \cr} 
  \nonumber\\
    && + \frac{\cos\theta}{\tan\beta} \pmatrix{
          d\, s\, b  \cr}
        \pmatrix{
          -m_u\e^{-i\delta_1-i\delta_2}\sin\alpha_1 &
          m_t x_d\e^{i\phi_{1d}} &
          m_u\e^{-i\delta_1-i\delta_2-i\delta_3}\cos\alpha_1\sin\alpha_2  \cr
          m_c\e^{-i\delta_1+i\delta_2}\cos\alpha_1 &
          m_t y_d\e^{i\phi_{2d}} &
          m_c\e^{-i\delta_1+i\delta_2-i\delta_3}\sin\alpha_1\sin\alpha_2  \cr
          0 &
          m_t z\e^{i\phi_3} &
          m_t\e^{2i\delta_1}\cos\alpha_2  \cr}
        \pmatrix{
          \bar{d} \cr  \bar{s} \cr  \bar{b}  \cr}
  \nonumber\\
    && + \frac{\cos\theta}{\tan\beta} \pmatrix{
          \bar{e}\, \bar{\mu}\, \bar{\tau}  \cr}
        \pmatrix{
          -m_u\e^{-i\delta_1-i\delta_2}\sin\alpha_1 &
          m_t x_l\e^{i\phi_{1l}} &
          m_u\e^{-i\delta_1-i\delta_2-i\delta_3}\cos\alpha_1\sin\alpha_2  \cr
          m_c\e^{-i\delta_1+i\delta_2}\cos\alpha_1 &
          m_t y_l\e^{i\phi_{2l}} &
          m_c\e^{-i\delta_1+i\delta_2-i\delta_3}\sin\alpha_1\sin\alpha_2  \cr
          0 &
          m_t z\e^{i\phi_3} &
          m_t\e^{2i\delta_1}\cos\alpha_2  \cr}
        \pmatrix{
          e \cr  \mu \cr  \tau  \cr}
  \nonumber\\
    && + \pmatrix{
          \bar{\nu_1}\, \bar{\nu_2}\, \bar{\nu_3}  \cr}
        \pmatrix{
          -m_u\e^{-i\delta_1-i\delta_2}\sin\alpha_1 &
          m_t \delta_x\e^{i\varphi_1} &
          m_u\e^{-i\delta_1-i\delta_2-i\delta_3}\cos\alpha_1\sin\alpha_2  \cr
          m_c\e^{-i\delta_1+i\delta_2}\cos\alpha_1 &
          m_t \delta_y\e^{i\varphi_2} &
          m_c\e^{-i\delta_1+i\delta_2-i\delta_3}\sin\alpha_1\sin\alpha_2  \cr
          0 &
          m_t \delta_z\e^{i\varphi_3} &
          m_t\e^{2i\delta_1}\cos\alpha_2  \cr}
        \pmatrix{
          \nu_1 \cr  \nu_2 \cr  \nu_3  \cr}
  \nonumber\\
    && + \frac{1}{2} \pmatrix{
          \bar{\nu_1}\, \bar{\nu_2}\, \bar{\nu_3}  \cr}
        \pmatrix{
          M_R\,\tilde{\jmath}_{11} & M_R\,\tilde{\jmath}_{12} & 
               M_R\,\tilde{\jmath}_{13}  \cr
          M_R\,\tilde{\jmath}_{21} & M_R\,\tilde{\jmath}_{22} & 
               M_R\,\tilde{\jmath}_{23}  \cr
          M_R\,\tilde{\jmath}_{31} & M_R\,\tilde{\jmath}_{32} & 
               M_R\,\tilde{\jmath}_{33}  \cr}
        \pmatrix{
          \bar{\nu_1} \cr  \bar{\nu_2} \cr  \bar{\nu_3}  \cr}, 
\label{mass_3}
\end{eqnarray}
in terms of the SM fields.
Here, we have defined various parameters as
\begin{eqnarray}
  x \equiv \frac{g_1}{h_{3}}\tan\theta, \qquad 
  y \equiv \frac{g_2}{h_{3}}\tan\theta, \qquad
  z \equiv \frac{g_3}{h_{3}}\tan\theta,
\end{eqnarray}
\begin{eqnarray}
  \delta_x \equiv \frac{k_1}{h_{3}}\frac{V}{M_G}, \qquad 
  \delta_y \equiv \frac{k_2}{h_{3}}\frac{V}{M_G}, \qquad
  \delta_z \equiv \frac{k_3}{h_{3}}\frac{V}{M_G},
\end{eqnarray}
and taken the effects of nonrenormalizable superpotential
Eq.~(\ref{GJ-superpot}) into account by changing $(1,2)$ and $(2,2)$
components of the down-type quark and charged-lepton mass matrices from
their original values $x\e^{i\phi_{1}}$ and $y\e^{i\phi_{2}}$.

Now, we redefine various SM fields as
\begin{eqnarray}
  \left\{ \begin{array}{l} 
    (u, d) \rightarrow (u, d)\,\e^{i\delta_1+i\delta_2} \\
    (c, s) \rightarrow (c, s)\,\e^{i\delta_1-i\delta_2} \\
    (t, b) \rightarrow (t, b)\,\e^{-2i\delta_1-i\delta_3}, \\
  \end{array} \right. \qquad
  \left\{ \begin{array}{l}
    \bar{u} \rightarrow \bar{u}\,\e^{-i\delta_1-i\delta_2} \\
    \bar{c} \rightarrow \bar{c}\,\e^{-i\delta_1+i\delta_2} \\
    \bar{t} \rightarrow \bar{t}\,\e^{2i\delta_1+i\delta_3}, \\
  \end{array} \right. \qquad
  \left\{ \begin{array}{l}
    \bar{d} \rightarrow \bar{d} \\
    \bar{s} \rightarrow \bar{s}\,\e^{-i\phi_3+2i\delta_1+i\delta_3} \\
    \bar{b} \rightarrow \bar{b}\,\e^{i\delta_3}, \\
  \end{array} \right.
\end{eqnarray}
\begin{eqnarray}
  \left\{ \begin{array}{l}
    (\nu_1, e)    \rightarrow (\nu_1, e)   \,\e^{i\delta_1+i\delta_2} \\
    (\nu_2, \mu)  \rightarrow (\nu_2, \mu) \,\e^{i\delta_1-i\delta_2} \\
    (\nu_3, \tau) \rightarrow (\nu_3, \tau)\,\e^{-2i\delta_1-i\delta_3}, \\
  \end{array} \right. \qquad
  \left\{ \begin{array}{l}
    \bar{\nu_1} \rightarrow \bar{\nu_1} \\
    \bar{\nu_2} \rightarrow \bar{\nu_2} \\
    \bar{\nu_3} \rightarrow \bar{\nu_3}, \\
  \end{array} \right. \qquad
  \left\{ \begin{array}{l}
    \bar{e}    \rightarrow \bar{e} \\
    \bar{\mu}  \rightarrow \bar{\mu} \,\e^{-i\phi_3+2i\delta_1+i\delta_3} \\
    \bar{\tau} \rightarrow \bar{\tau}\,\e^{i\delta_3}, \\
  \end{array} \right.
\end{eqnarray}
in order to simplify the form of the quark and lepton mass matrices.
Then, from Eq.~(\ref{mass_3}) we obtain the mass matrices defined by
\begin{eqnarray}
  W &=& \pmatrix{
          u\, c\, t  \cr}
        M_u
        \pmatrix{
          \bar{u} \cr  \bar{c} \cr  \bar{t}  \cr} 
       + \pmatrix{
          d\, s\, b  \cr}
        M_d
        \pmatrix{
          \bar{d} \cr  \bar{s} \cr  \bar{b}  \cr}
       + \pmatrix{
          \bar{e}\, \bar{\mu}\, \bar{\tau}  \cr}
        M_l
        \pmatrix{
          e \cr  \mu \cr  \tau  \cr}
  \nonumber\\
    && + \pmatrix{
          \bar{\nu_1}\, \bar{\nu_2}\, \bar{\nu_3}  \cr}
        M_{\nu D}
        \pmatrix{
          \nu_1 \cr  \nu_2 \cr  \nu_3  \cr}
       + \frac{1}{2} \pmatrix{
          \bar{\nu_1}\, \bar{\nu_2}\, \bar{\nu_3}  \cr}
        M_N
        \pmatrix{
          \bar{\nu_1} \cr  \bar{\nu_2} \cr  \bar{\nu_3}  \cr}, 
\end{eqnarray}
as
\begin{eqnarray}
  M_u &=& m_t \pmatrix{
          \hat{m}_u & 0 & 0  \cr
          0 & \hat{m}_c & 0  \cr
          0 & 0 & 1  \cr},
  \nonumber\\
  M_d &=& m_t \frac{\cos\theta}{\tan\beta} \pmatrix{
          -\hat{m}_u\sin\alpha_1 &
          x_d\e^{i\phi_{xd}} &
          \hat{m}_u\cos\alpha_1\sin\alpha_2  \cr
          \hat{m}_c\cos\alpha_1 &
          y_d\e^{i\phi_{yd}} &
          \hat{m}_c\sin\alpha_1\sin\alpha_2  \cr
          0 &
          z &
          \cos\alpha_2  \cr},
  \nonumber\\
  M_l &=& m_t \frac{\cos\theta}{\tan\beta} \pmatrix{
          -\hat{m}_u\sin\alpha_1 &
          x_l\e^{i\phi_{xl}} &
          \hat{m}_u\cos\alpha_1\sin\alpha_2  \cr
          \hat{m}_c\cos\alpha_1 &
          y_l\e^{i\phi_{yl}} &
          \hat{m}_c\sin\alpha_1\sin\alpha_2  \cr
          0 &
          z &
          \cos\alpha_2  \cr},
  \nonumber\\
  M_{\nu D} &=& m_t \pmatrix{
          -\hat{m}_u\sin\alpha_1 &
          \delta_x\e^{i\varphi_x} &
          \hat{m}_u\e^{i\varphi_u}\cos\alpha_1\sin\alpha_2  \cr
          \hat{m}_c\e^{i\varphi_t}\cos\alpha_1 &
          \delta_y\e^{i\varphi_y} &
          \hat{m}_c\e^{i\varphi_v}\sin\alpha_1\sin\alpha_2  \cr
          0 &
          \delta_z\e^{i\varphi_z} &
          \e^{i\varphi_w}\cos\alpha_2  \cr},
  \nonumber\\
  M_N &=& M_R \pmatrix{
          \tilde{\jmath}_{11} & \tilde{\jmath}_{12} & 
               \tilde{\jmath}_{13}  \cr
          \tilde{\jmath}_{21} & \tilde{\jmath}_{22} & 
               \tilde{\jmath}_{23}  \cr
          \tilde{\jmath}_{31} & \tilde{\jmath}_{32} & 
               \tilde{\jmath}_{33}  \cr},
\end{eqnarray}
where $\hat{m}_u \equiv m_u/m_t,\, \hat{m}_c \equiv m_c/m_t$.
Here, we have defined various phases as
\begin{eqnarray}
  \matrix{
\phi_{xd} \equiv \phi_{1d} - \phi_3 + 3\delta_1 + \delta_2 + \delta_3, \qquad
\phi_{yd} \equiv \phi_{2d} - \phi_3 + 3\delta_1 - \delta_2 + \delta_3, \cr
\phi_{xl} \equiv \phi_{1l} - \phi_3 + 3\delta_1 + \delta_2 + \delta_3, \qquad
\phi_{yl} \equiv \phi_{2l} - \phi_3 + 3\delta_1 - \delta_2 + \delta_3, \cr}
\end{eqnarray}
\begin{eqnarray}
  \matrix{
\varphi_x \equiv \varphi_1 + \delta_1 - \delta_2, \qquad
\varphi_y \equiv \varphi_2 + \delta_1 - \delta_2, \qquad
\varphi_z \equiv \varphi_3 + \delta_1 - \delta_2, \cr
\varphi_t \equiv   2\delta_2, \qquad
\varphi_u \equiv - 3\delta_1 - \delta_2 - 2\delta_3, \cr
\varphi_v \equiv - 3\delta_1 + \delta_2 - 2\delta_3, \qquad
\varphi_w \equiv -  \delta_3. \cr}
\end{eqnarray}
Note that we have retained the form of $M_N$.
As a result, not all phases are physically independent.
The domains of $\alpha_1$ and $\alpha_2$ can be restricted to
\begin{eqnarray}
  0 \leq \alpha_1 \leq \pi/2, \qquad
  0 \leq \alpha_2 \leq \pi/2,
\end{eqnarray}
without a loss of generality by redefining the phases of the SM fields
appropriately.

Finally, the mass matrix for the light neutrinos defined by
\begin{eqnarray}
  W &=& \frac{1}{2} \pmatrix{
          \nu_1\, \nu_2\, \nu_3  \cr}
        M_{\nu}
        \pmatrix{
          \nu_1 \cr  \nu_2 \cr  \nu_3  \cr}, 
\end{eqnarray}
is given by
\begin{eqnarray}
  M_{\nu} = M_{\nu D}^{\rm T} M_N^{-1} M_{\nu D},
\end{eqnarray}
through see-saw mechanism \cite{Seesaw}.

\newpage
%
%%%%%%%%%%%%%%%%%%%%%%%%%%%%%%%%%%%%%%%%%%%%%%%%%%%%%%%%%%%%%%%
%
% NEW COMMANDS FOR THE BIBLIOGRAPHY
%
%%%%%%%%%%%%%%%%%%%%%%%%%%%%%%%%%%%%%%%%%%%%%%%%%%%%%%%%%%%%%%%
\newcommand{\Journal}[4]{{\sl #1} {\bf #2} {(#3)} {#4}}
\newcommand{\PL}{\sl Phys. Lett.}
\newcommand{\PR}{\sl Phys. Rev.}
\newcommand{\PRL}{\sl Phys. Rev. Lett.}
\newcommand{\NP}{\sl Nucl. Phys.}
\newcommand{\ZP}{\sl Z. Phys.}
\newcommand{\PTP}{\sl Prog. Theor. Phys.}
\newcommand{\NC}{\sl Nuovo Cimento}
\newcommand{\MPL}{\sl Mod. Phys. Lett.}
\newcommand{\PRep}{\sl Phys. Rep.}
%%%%%%%%%%%%%%%%%%%%%%%%%%%%%%%%%%%%%%%%%%%%%%%%%%%%%%%%%%%%%%%

%

\end{document}